\documentclass[12pt]{article}
 \usepackage{hyperref}
 \usepackage{cite}
 \usepackage{graphicx}
  \usepackage{epsfig}
  \usepackage{amssymb}
  \usepackage{subfigure}
  \usepackage{amsmath,color}

\evensidemargin - .5truecm
\allowdisplaybreaks[1]

\begin{document}

\newcommand{\lsim}{\lesssim}
\newcommand{\gsim}{\gtrsim}

\newcommand{\CC}{{\mathbb C}}
\newcommand{\RR}{{\mathbb R}}
\newcommand{\ZZ}{{\mathbb Z}}
\newcommand{\QQ}{{\mathbb Q}}
\newcommand{\NN}{{\mathbb N}}
\newcommand{\beq}{\begin{equation}}
\newcommand{\eeq}{\end{equation}}
\newcommand{\beal}{\begin{align}}
\newcommand{\eeal}{\end{align}}
\newcommand{\nn}{\nonumber}
\newcommand{\bea}{\begin{eqnarray}}
\newcommand{\eea}{\end{eqnarray}}
\newcommand{\ba}{\begin{array}}
\newcommand{\ea}{\end{array}}
\newcommand{\bfig}{\begin{figure}}
\newcommand{\efig}{\end{figure}}
\newcommand{\bc}{\begin{center}}
\newcommand{\ec}{\end{center}}

\newenvironment{appendletterA}
{
  \typeout{ Starting Appendix \thesection }
  \setcounter{section}{0}
  \setcounter{equation}{0}
  \renewcommand{\theequation}{A\arabic{equation}}
 }{
  \typeout{Appendix done}
 }
\newenvironment{appendletterB}
 {
  \typeout{ Starting Appendix \thesection }
  \setcounter{equation}{0}
  \renewcommand{\theequation}{B\arabic{equation}}
 }{
  \typeout{Appendix done}
 }

\begin{titlepage}
\nopagebreak

\renewcommand{\thefootnote}{\fnsymbol{footnote}}
\vskip 1cm
%\begin{center}
%\boldmath

\vspace*{.5cm}
\begin{center}
{\Large \bf
Factorization
of the Energy-Energy Correlation \\ 
in the two-jet limit
in the massive case
\\[0.1cm]
}
\end{center}

%{\Large\bf

\par \vspace{1.5mm}
\begin{center}
  {\bf  Ugo Giuseppe\ Aglietti${}^{(a)}$,~}
  {\bf Giancarlo Ferrera${}^{(b)}$}
  and~{\bf Lorenzo Rossi${}^{(b)}$  
 }\\

\vspace{5mm}

${}^{(a)}$
Dipartimento di Fisica, Universit\`a di Roma ``La Sapienza'' and\\ INFN, Sezione di Roma
I-00185 Rome, Italy\\\vspace{1mm}

${}^{(b)}$
Dipartimento di Fisica, Universit\`a di Milano and\\ INFN, Sezione di Milano,
I-20133 Milan, Italy\\\vspace{1mm}

\end{center}

\vspace{.5cm}

%\pacs{13.87.Ce,12.38Bx}

\par %\vspace{2mm}
\begin{center} {\large \bf Abstract} \end{center}
\begin{quote}
\pretolerance 10000
%\begin{abstract}

We consider non-logarithmic heavy-quark mass effects
in the factorization and resummation
of the Energy-Energy-Correlation (EEC) function,
in the two-jet limit.
{\color{black} We define a new partial EEC cumulant, normalized at a fixed
correlation angle $\chi_M<\pi$. At first order, the back-to-back
endpoint coefficient cancels in this ratio, while the forward endpoint
does not enter. The result can therefore be obtained from the
real-emission EEC distribution in $D=4$ space-time dimensions, without
determining the endpoint contributions separately.}
In order to determine explicitly the
next-to-leading order
coefficient function and the remainder function
(both entering the
standard resummation formula),
we evaluate numerically the
EEC spectrum at first order in
$\alpha_S$, finding good agreement
with previous calculations.
To have a smooth massless
limit, a new, improved factorization
scheme is proposed, in which
the coefficient function also
depends on the correlation angle $\chi$.

%\end{abstract}
\vfill
\end{quote}
\vspace*{0.5cm}
\vspace*{1cm}
\vspace*{\fill}

\begin{flushleft}
July 2026
\vspace*{-1cm}
\end{flushleft}
\end{titlepage}

\renewcommand{\thefootnote}{\fnsymbol{footnote}}

\newpage

\tableofcontents

\newpage

%%%%%%%%%%%%%%%%%%%%%%

\section{Introduction}

Given the high accuracy currently achieved in perturbative QCD (pQCD)
calculations of the Energy-Energy Correlation
(EEC) function \cite{Basham:1978bw},
fixed-order
\cite{Richards:1982te, Richards:1983sr, Dixon:2018qgp,
DelDuca:2016csb, DelDuca:2016ily, Moult:2018jzp, Ebert:2020sfi, Duhr:2022yyp}
as well as resummed ones
\cite{Collins:1981uk, Collins:1981va,
Collins:1985xx, Kodaira:1981nh,
Kodaira:1982cr, Kodaira:1982az,
deFlorian:2004mp, Tulipant:2017ybb,
Kardos:2018kqj, Dixon:2019uzg, Aglietti:2024xwv},
the inclusion of (perturbative) heavy-quark
mass effects, at least at leading order
in $\alpha_S$, is strongly recommended for high-precision phenomenology, as also observed
in other observables~\cite{Aglietti:2006wh,Aglietti:2007bp,Aglietti:2008xn,Aglietti:2022rcm,Caletti:2023spr,Ghira:2023bxr,Gaggero:2022hmv,Caletti:2022hnc,Fedkevych:2022mid,DiGiustino:2007mba,Catani:2000ef,Catani:2002hc,Cacciari:2002re,Corcella:2003ib,Ghira:2024nkk,vonKuk:2023jfd,vonKuk:2024uxe}.

The EEC differential distribution
(or spectrum) is defined as~\cite{Basham:1978bw}:
\beq
\frac{d\Sigma}{d\chi}
\, \equiv \,
\sum_{n=2}^\infty
\sum_{i,j=1}^n
\int
\frac{E_i E_j}{Q^2} \,
d\sigma^{e^+e^-\to ijX}
\delta\left(\cos\chi+\cos\theta_{ij}\right);
\eeq
where $Q\equiv \sqrt{s}$ is the hard scale
of the process, $\theta_{ij}$ is the
(unoriented) angle between the three (space)
momenta of hadrons (partons in pQCD) $i$ and $j$
($0 \le \theta_{ij}\le \pi$). The variable $\chi$ denotes the correlation angle,
\beq
\chi \, = \, \pi \, - \, \theta_{ij},
\eeq
vanishing in the two-jet limit
(back-to-back pairs)
and having $\pi$ as the limit in the forward region
(collinear pairs).
The EEC function thus describes the distribution
of the angular separation between the hard
particle pairs.

In a previous work \cite{Aglietti:2024zhg}, two of the present authors
considered {\it logarithmic} heavy-quark mass effects,
which have been factorized
and resummed to all orders in $\alpha_S$
in a specific (massive)
Sudakov form factor, entering the standard
factorization formula.
In this paper, we concentrate instead 
on {\it non-logarithmic} mass effects,
entering only the coefficient function
and the remainder function.

As is well known, at tree level the EEC spectrum has,
both in the massless and in the massive case,
two peaks of equal strength at its endpoints
$\chi=0$ and $\chi=\pi$:
\beq
\frac{1}{\sigma_0} \,
\frac{d\Sigma}{d\chi}
\, = \,
\frac{1}{2} \, \delta(\chi) \, + \,
\frac{1}{2} \, \delta(\pi-\chi)
\, + \, \mathcal{O}\left(\alpha_S\right),
\eeq
where $\sigma_0$ is the total
cross section in Born approximation.
{\color{black} Beyond the Born approximation, the virtual correction to
$e^+e^-\mapsto Q\overline Q$ retains the two-body kinematics and
therefore contributes equally to the two endpoint delta functions.  The
massive one-loop quark form factors determining this common virtual term
are known.  The complete endpoint coefficients, however, also contain
distributional contributions from the real process
$e^+e^-\mapsto Q\overline Qg$, including the self-correlations at the
forward endpoint.  Their separate determination requires a dedicated and
consistent treatment of the real and virtual endpoint contributions.

In sec.$\,$\ref{sec_new_EF}, we define a partial EEC cumulant,
obtained by normalizing the EEC cumulant at a fixed angle
$\chi_M<\pi$. At first order, the back-to-back endpoint coefficient
cancels from this ratio, while the forward endpoint does not enter. The
quantity can therefore be evaluated from the real-emission EEC
distribution in four space-time dimensions.}

In sec.$\,$\ref{sec_EEC_FO} we explicitly evaluate the EEC spectrum
for a heavy quark ($m \ne 0$) to first order
in $\alpha_S$.
In general, a non-zero quark mass introduces
radicals in place of simple powers
of the energy or of the space momentum
of the quark; this renders the computation
of the EEC distribution much harder
than in the massless case.
As we are going to show in detail,
the EEC cumulant
can be written as a two-dimensional
integral over the quark and antiquark
energies.
We have been able to make analytically
the first integration;
for the second integration,
we had to resort to numerical methods.
Our results for the EEC spectrum are then given in terms of
a one-dimensional integral, depending
parametrically on the correlation
angle $\chi$ and the mass
correction parameter $\eta \equiv 2m/Q$
$(0 \le \eta \le 1)$, where $m$ is the
heavy-quark mass.
The first-order calculation of the massive EEC
function was performed numerically some time ago in
\cite{Csikor:1983dt,Ali:1984gzn};
we find good agreement with both these papers.

To be more specific, in sec.$\,$\ref{sec_Qg}, 
we evaluate the quark-gluon
(or antiquark-gluon)
correlation.
We give one-dimensional integral representations of both the differential and cumulative distributions.
In sec.$\,$\ref{sec_QQbar},
we consider the quark-antiquark correlation.
The latter is more complicated than
the quark-gluon correlation because it involves
two massive partons rather than one;
roughly speaking, ``more nonlinearity''
is involved.

In sec.$\,$\ref{sec_stand_fact_scheme}
we extend in a straightforward way the standard massless
factorization scheme to the massive case.
We evaluate (numerically) the
leading-order massive remainder functions,
for a wide range of the mass correction
parameter $\eta$ ($0 \le \eta \le 0.9$).
We find that massive remainder functions
can be related with a good accuracy in a simple way
to the simpler massless
remainder function, so long as the quark mass
parameter is not too large ($\eta \le 0.5$).

The standard massive factorization scheme
turns out to be not smooth
(continuous) in the massless limit.
That is to say that the massive remainder
function
and the massive coefficient function do not tend
to the corresponding {\it ab initio} massless
quantities in the limit $\eta \to 0^+$.
That is not a serious issue as long as one
considers very heavy quarks, but
is unpleasant for small masses.

In sec.$\,$\ref{sec_new_massive_factor}
we then construct an improved factorization
scheme, having the desired smooth massless limit.
To that aim, we are forced to introduce
a dependence,
in the new coefficient function,
 on the correlation angle
$\chi$.
An analogous problem to the present one has been found
a few years ago in threshold
resummation in the massive case
\cite{Aglietti:2022rcm,Gaggero:2022hmv,Ghira:2023bxr,Ghira:2024nkk}, where a similar solution has been found.

Finally, Sec.~\ref{sec_conclude} presents our conclusions and
outlines several possible developments.
In particular, improved theoretical control over heavy-quark mass effects opens up new possibilities for precision phenomenological analyses.

{\color{black} In order to make the paper reasonably self-contained, we have also added two appendices.
In Appendix~\ref{sec_kino} we consider in some depth $Q\overline{Q}g$
kinematics, namely the 3-body kinematics
with a final massive quark-antiquark pair
and a (massless) gluon.
In Appendix~\ref{sec_ME} we report the matrix elements
squared
for one gluon emission off a massive
$Q\overline{Q}$ pair \cite{Dokshitzer:1994jt},
which generalize
the well-known formula for the massless
case.
We consider both $e^+e^-$ annihilation
into a photon and a $Z^0$, i.e. the cases
of a $Q\overline{Q}$ vector and axial currents.}

%%%%%%%%%%%%%%%%%%%%%%%%%
{\color{black}
\section{EEC cumulants}
\label{sec_new_EF}

In this section we first recall the standard normalized EEC cumulant
and then introduce a partial EEC cumulant normalized at a fixed
correlation angle $\chi_M$.  We begin with a simpler example in order
to illustrate the cancellation of endpoint contributions in normalized
cumulative distributions.

%%%%%%%%%%%%%%%%%%%%%%%%%%%%%%%%%%%%

\subsection{Standard EEC cumulant}

The standard normalized EEC cumulant is defined as:}
\beq
\label{eq_tot_ev_frac}
R\left( \chi; \, \alpha_S\right)
\, \equiv \,
\frac{ \int\limits_0^{\chi}
d\Sigma/d\chi'\left( \chi'; \, \alpha_S \right) \, d\chi' }{ \int\limits_0^{\pi} d\Sigma/d\chi'\left( \chi'; \, \alpha_S \right) \, d\chi' };
\qquad
0 \le \chi \le \pi.
\eeq
It has the following properties:
%%%%%%%%%%%%%%%%%
\begin{enumerate}
%%%%%
\item
It vanishes in the two-jet (or back-to-back) limit,
because of Sudakov suppression:
\beq
\lim_{\chi\to 0^+}
R\left( \chi; \, \alpha_S\right)
\, = \, 0.
\eeq
We recall that fixed-order perturbation theory
does not satisfy the above condition.
The two-jet limit involves indeed typical Sudakov
effects, so the above condition is satisfied by resummed perturbative calculations only.
%%%%%
\item
By unitarity, it equals unity
in the forward limit,
since the cumulant covers the full angular range:
\beq
\lim_{\chi\to \pi^-}
R\left( \chi; \, \alpha_S\right)
\, = \, 1.
\eeq
%%%%%
\item
The EEC cumulant is a monotonically increasing
function of $\chi$, as its
derivative is a (differential) cross
section, which is a positive-definite
quantity.
%%%%%
\item
The differential distribution,
or spectrum --- which is the quantity usually
measured in experiments ---
is trivially obtained
by differentiating the cumulant
with respect to
$\chi$.
%%%%%%%%%%%%%%%
\end{enumerate}

%%%%%%%%%%%%%%%%%%%%%%%%%%%%%%%%%%%%%%%%%%%%%%%%%%%%

\subsubsection{A simple example from beauty physics}

To begin with,
let us see what the event-fraction looks
like in a simpler case than the EEC,
namely the invariant mass distribution in $B \to X_s \gamma$ decays, i.e.\ the spectrum in
\begin{equation}
y \equiv \frac{m^2_{X_s}}{m^2_B} \qquad (m_s = 0),
\end{equation}
where $m_B$ is the $B$-meson mass, $m_{X_s}$ is the invariant mass of the hadronic 
recoil system $X_s$, and $m_s = 0$ indicates that the strange quark mass is neglected.
By introducing an infrared regulator
(such as Dimensional Regularization or
a small gluon mass), the first-order
spectrum can be written schematically:
\beq
\frac{1}{\Gamma} \, \frac{d\Gamma}{dy}
\, = \, \left(1 \, + \, C_1 \, \alpha_S\right) \, \delta(y)
\, + \, \alpha_S \, f(y)
\, + \, \mathcal{O}\left(\alpha_S^2\right)
\qquad (0 \le y \le 1).
\eeq
The first-order expansion of the event fraction
then reads:
\bea
R\left( y; \alpha_S \right)
&\equiv& \frac{
\int\limits_0^y d\Gamma/dy' \, dy'}{
\int\limits_0^1 d\Gamma/dy' \, dy'}
\, = \, \frac{1 \, + \, C _1 \, \alpha_S
\, + \, \alpha_S
\int\limits_0^y f(y') \, dy'
\, + \, \mathcal{O}(\alpha_S^2)}{
1 \, + \, C_1 \, \alpha_S \, + \, \alpha_S
\int\limits_0^1 f(y') \, dy'
\, + \, \mathcal{O}(\alpha_S^2)} \, =
\nonumber\\
&=& 1 \, - \, \alpha_S \int\limits_y^1 f(y') \, dy'
\, + \, \mathcal{O}(\alpha_S^2).
\eea
Let us comment on the above result.
The one-loop virtual corrections,
represented by the $C_1$ term,
cancel in the ratio, so they do not
contribute to the event fraction.
That is in agreement with physical intuition:
one-loop corrections have the trivial
Born kinematics and then are not expected
to give a contribution to quantities
describing some structure
of the final hadronic states.
The idea behind the event fraction is indeed
that of separating the (total) cross section
of a class of events from the
cross sections describing the distribution
of such events.
We normalize indeed the distribution to the total
radiatively-corrected cross section 
$\sigma=\sigma(\alpha_S)$,
rather than to the Born cross section $\sigma_0$.

%%%%%%%%%%%%%%%%%%%%%%%%%%%%%%%%%%%%%%%%%%%%%%%%%

\subsubsection{Perturbative expansion in the EEC case}

The main complication of the EEC
with respect to simpler cases
such as the one described in the previous section,
comes
from the fact that its lowest-order
spectrum contains,
as already remarked, two peaks, one at each endpoint, rather than just one.
At first order, one can write:
\beq
\frac{1}{\sigma} \frac{d\Sigma}{d\chi}
\, = \,
\frac{1}{2} \left(1+C_1\alpha_S\right) \, \delta(\chi)
\, + \, \alpha_S \, f(\chi)
\, + \, \frac{1}{2} \left(1+K_1\alpha_S\right)
\, \delta(\chi-\pi)
\, + \, \mathcal{O}\left(\alpha_S^2\right).
\eeq
The first-order expansion of the standard EEC cumulant then reads:
\beq
\label{eq_EEC_EF_no}
R(\chi;\,\alpha_S)
\, = \, \frac{1}{2}
\left[
1 \, + \, \frac{1}{2} \left(C_1-K_1\right) \alpha_S
\, + \, \alpha_S \int\limits_0^{\chi}
f(\chi') \, d \chi'
\, - \, \alpha_S \int\limits_{\chi}^\pi
f(\chi') \, d \chi'
\right] \, + \, \mathcal{O}\left(\alpha_S^2\right).
\eeq
{\color{black} The coefficients $C_1$ and $K_1$ are the complete endpoint
coefficients of the normalized EEC distribution. The virtual correction
retains Born kinematics and, together with the common contribution induced
by the normalization to the total cross section, enters both coefficients
in the same way. Their difference is therefore generated by the
real-emission endpoint terms. In particular, the self-correlations of the
$Q\overline Qg$ final state contribute at the forward endpoint. Thus, the
massive quark form factor alone does not determine $C_1$ and $K_1$; their
separate calculation requires a consistent distributional treatment of
the real-emission endpoint contributions. It is therefore useful to
introduce a quantity in which the back-to-back endpoint coefficient
cancels and the forward endpoint does not enter.}

%%%%%%%%%%%%%%%%%%%%%%%%%%%%%%%%%%%%%%%%%%%%

\subsection{{\color{black}A new partial EEC cumulant}}

Let us then consider the quantity
\beq
\label{eq_part_ev_frac}
R\left( \chi; \, \chi_M ; \, \alpha_S\right)
\, \equiv \,
\frac{ \int\limits_0^{\chi}
d\Sigma/d\chi'\left( \chi'; \, \alpha_S \right) \, d\chi' }{ \int\limits_0^{\chi_M} d\Sigma/d\chi'\left( \chi'; \, \alpha_S \right) \, d\chi' },
\eeq
with:
\beq
0 \, < \, \chi_M \, < \, \pi.
\eeq
Typically, one takes:
\beq
\chi_M \, \sim \, \frac{\pi}{2}.
\eeq
In a given analysis, the parameter
$\chi_M$ is fixed, while
$\chi$ is a variable with the range
\beq
0 \, \le \, \chi \, \le \, \chi_M.
\eeq
The standard EEC cumulant is formally recovered for $\chi_M=\pi$.
In the following, however, $\chi_M=\mathcal O(1)$ is kept fixed and
away from the forward endpoint.

The partial EEC cumulant has similar
properties to the old one
(see eq.(\ref{eq_tot_ev_frac})):
%%%%%%%%%%%%%%%%%
\begin{enumerate}
%%%%%
\item
  It vanishes in the two-jet limit
    because of Sudakov suppression:
\beq
\lim_{\chi\to 0^+}
R\left( \chi; \, \chi_M ; \, \alpha_S\right)
\, = \, 0.
\eeq
%%%%%
\item
By definition:
\beq
\label{eq_new_EF_unitarity}
R\left( \chi = \chi_M;
\, \chi_M ; \, \alpha_S \right)
\, \equiv \, 1.
\eeq
%%%%%
\item
The corresponding normalized differential distribution
is obtained again by differentiating
the new partial EEC cumulant
with respect to
$\chi$.
%%%%%%%%%%%%%%%
\end{enumerate}
{\color{black} Equation~(\ref{eq_part_ev_frac}) does not define a restricted
sample of events.  The EEC is a pairwise correlation, and a
higher-multiplicity event generally contributes through several pairs with
different values of $\chi$, possibly both below and above $\chi_M$.
The quantity in Eq.~(\ref{eq_part_ev_frac}) is simply the ratio of two
integrals of the same inclusive, infrared- and collinear-safe EEC
distribution.

The upper integration limit therefore does not impose a veto on radiation
and does not introduce measured and unmeasured angular regions in the sense
relevant for non-global logarithms.  For fixed $\chi_M=\mathcal O(1)$,
away from the endpoints, the denominator is a finite normalization factor
and introduces no additional small scale.  Hence the logarithmically
enhanced dependence for $\chi\to0$ is described by the standard
back-to-back EEC Sudakov factor, with no additional non-global evolution.

Let us note that, on the r.h.s. of
eq.(\ref{eq_part_ev_frac}), one could replace $d\Sigma/d\chi$ for example
with $1/\sigma \, d\Sigma/d\chi$,
as overall factors cancel between
the numerator and the denominator. The same differential distribution
must be used in the numerator and denominator of
Eq.~(\ref{eq_part_ev_frac}).}

%%%%%%%%%%%%%%%%%%%%%%%%%%%%%%%%%%%%%%%%%%%%%%%

\subsubsection{Fixed-order perturbation theory}

Unlike Eq.~(\ref{eq_EEC_EF_no}), the first-order expansion of the
partial EEC cumulant reads:
\beq
R\left( \chi; \, \chi_M ; \, \alpha_S\right)
\, = \, 1 \, - \, 2 \, \alpha_S
\int\limits_{\chi}^{\chi_M}
f(\chi') \, d \chi'
\, + \, \mathcal{O}\left(\alpha_S^2\right).
\eeq
The back-to-back endpoint coefficient cancels between the numerator
and the denominator, while the forward endpoint does not enter, and a
simple expression is obtained.
The factor two in front of the
$\mathcal{O}(\alpha_S)$ correction term
originates from the fact that we only
integrate, both at the numerator and the
denominator, over the two-jet or back-to-back peak,
which has a strength $=1/2$.

%%%%%%%%%%%%%%%%%%%%%%%%%%%%%%%%%%%%%%%%%%%%

\subsubsection{Resummed perturbation theory}

Let us now consider the partial EEC cumulant in resummed perturbation theory.
As is well known, the differential EEC cross section is written in factorized form as~\cite{Catani:1992ua}:
\beq
\label{eq_EEC_diff_fact}
\frac{1}{\sigma} \,
\frac{d\Sigma}{d\chi}\left( \chi; \alpha_S \right)
\, = \, \frac{1}{2} H(\alpha_S) \, \varphi\left( \chi; \alpha_S \right)
\, + \, \frac{C_F \alpha_S}{2\pi} \, \mathrm{res}\left( \chi; \alpha_S \right);
\eeq
where $\sigma=\sigma(\alpha_S)$ is the radiatively-corrected
total cross section for
$e^+e^-\mapsto\mathrm{hadrons}$,
\beq
\sigma \, = \, \int\limits_0^\pi \frac{d\Sigma}{d\chi'} d\chi',
\eeq
and:
%%%%%%%%%%%%%%%%%
\begin{enumerate}
%%%%%
\item
$H(\alpha_S)$ is a hard-virtual factor,
having a standard perturbative expansion:
\bea
\label{eq_first_hard_factor}
H(\alpha_S) &=& 1 \, + \,
\frac{C_F}{2} \sum_{n=1}^\infty
\left( \frac{\alpha_S}{\pi} \right)^n H^{(n)}
\, =
\nonumber\\
&=& 1 \, + \, \frac{C_F \, \alpha_S}{2\pi} H^{(1)}
\, + \, \frac{C_F}{2}
\left( \frac{\alpha_S}{\pi} \right)^2 H^{(2)}
\, + \, \cdots.
\eea
In the perturbative corrections,
we have introduced an overall factor
$C_F/2$, as most higher-order
corrections are directly proportional
to $C_F$.
%%%%%
\item
$\varphi\left( \chi; \alpha_S \right)$
is a Sudakov form factor factorizing
and resumming large double logarithms
of infrared (soft and/or collinear) origin.
%%%%%
\item
$\mathrm{res}\left( \chi; \alpha_S \right)$
is a short-distance remainder function,
having a standard perturbative expansion:
\bea
\frac{C_F \alpha_S}{2\pi} \,
\mathrm{res}\left( \chi; \alpha_S \right)
&=&
\frac{C_F}{2} \sum_{n=1}^\infty
\left( \frac{\alpha_S}{\pi} \right)^n
\mathrm{res}^{(n)}\left( \chi \right) =
\nonumber\\
&=& \frac{C_F \alpha_S}{2\pi} \,
\mathrm{res}^{(1)}\left( \chi \right)
\, + \, \frac{C_F}{2}
\left( \frac{\alpha_S}{\pi} \right)^2
\mathrm{res}^{(2)}\left( \chi \right)
\, + \, \cdots.
\eea
To make subsequent perturbative expansions easier,
we have pulled out from the remainder
function a factor $C_F\alpha_S/(2\pi)$.
That is legitimate, as the remainder
function (unlike the hard factor
and the Sudakov form factor) does not
possess a $\mathcal{O}\left(\alpha_S^0\right)$ term.
%%%%%%%%%%%%%%%
\end{enumerate}
By formally integrating eq.(\ref{eq_EEC_diff_fact}) on both sides over
$\chi$, one obtains:
\beq
\frac{1}{\sigma(\alpha_S)} \,
\Sigma\left( \chi; \, \alpha_S \right)
\, \equiv \,
\frac{1}{\sigma(\alpha_S)}
\int\limits_0^{\chi}
\frac{d\Sigma}{d\chi'}\left( \chi'; \alpha_S \right) \, d \chi'
\, = \,
\frac{1}{2} H\left(\alpha_S\right) \, \Phi\left( \chi; \, \alpha_S \right)
\, + \, \frac{C_F \, \alpha_S}{2\pi} \, \mathrm{Res}\left( \chi; \, \alpha_S \right);
\eeq
where we have defined:
\bea
\Phi\left( \chi; \alpha_S \right)
&\equiv&
\int\limits_0^{\chi} \,\varphi\left( \chi'; \alpha_S \right) d \chi';
\nonumber\\
\mathrm{Res}\left( \chi; \alpha_S \right)
&\equiv&
\int\limits_0^{\chi} \mathrm{res}\left( \chi'; \, \alpha_S \right) \, d \chi'.
\eea
An important property of the partially-integrated
Remainder function is to vanish
in the two-jet limit:
\beq
\label{eq_imp_prop_Rem_fun}
\lim_{\chi\to 0^+}
\mathrm{Res}\left( \chi; \alpha_S \right)
\, = \, 0.
\eeq
In general, let us use capital letters for the
partially-integrated distributions.
The partial EEC cumulant is then written:
\beq
\label{eq_new_EF_interm}
R\left( \chi; \, \chi_M ; \, \alpha_S\right)
\, = \, \frac{\Phi\left( \chi; \, \alpha_S \right)
\, + \, C_F \, \alpha_S/\pi \, \mathrm{Rel}\left( \chi; \, \alpha_S \right)}{
\Phi_M
\, + \, C_F \, \alpha_S/\pi \,
\mathrm{Rel}_M
};
\eeq
where we have divided both numerator
and denominator by $1/2\,H(\alpha_S)$
and we have defined the new Remainder function:
\beq
\mathrm{Rel}\left( \chi; \, \alpha_S \right)
\, \equiv \,
\frac{\mathrm{Res}\left( \chi; \alpha_S \right)}{H(\alpha_S)}.
\eeq
We have also defined:
\bea
\Phi_M &\equiv& \Phi\left( \chi=\chi_M; \, \alpha_S \right);
\nonumber\\
\mathrm{Rel}_M &\equiv&
\mathrm{Rel}\left( \chi=\chi_M ; \, \alpha_S \right).
\eea
Up to first-order (i.e. up to the lowest non-vanishing, non-trivial order), the new and the old Remainder
functions are equal:
\beq
\mathrm{Rel}^{(1)}\left( \chi\right)
\, = \,
\mathrm{Res}^{(1)}\left( \chi\right).
\eeq
Note that, on the r.h.s. of eq.(\ref{eq_new_EF_interm}),
the new Remainder function
is multiplied by $C_F\alpha_S/\pi$,
rather than the previous $C_F\alpha_S/(2\pi)$.

\noindent
The partial EEC cumulant can then be written in factorized form as:
\beq
\label{eq_EEC_EF_main_eq}
R\left( \chi, \, \chi_M ; \, \alpha_S \right)
\, = \,
C(\alpha_S) \,
\frac{\Phi\left( \chi; \, \alpha_S \right)}{\Phi_M}
\, + \, \frac{C_F \alpha_S}{\pi} \,
\mathrm{Rem}\left( \chi ; \alpha_S\right);
\eeq
where we have defined the new coefficient
function:
\beq
C\left(\alpha_S \right)
\,\equiv \,
\frac{1}{1 \, + \, C_F \, \alpha_S/\pi \,
\mathrm{Rel}_M/\Phi_M }.
\eeq
Note that it also depends on $\chi_M.$
We have also defined the new Remainder function
\beq
\mathrm{Rem}\left( \chi ; \alpha_S\right)
\, \equiv \,
\frac{
\mathrm{Rel}\left( \chi ; \alpha_S\right)
}{
\Phi_M \, + \, C_F \, \alpha_S/\pi \, \mathrm{Rel}_M }.
\eeq
To simplify forthcoming formulae, it is convenient
to define the perturbative expansion
of the new coefficient function $C(\alpha_S)$
in a slightly different way with respect
to the old one $H(\alpha_S)$
(cf. eq.(\ref{eq_first_hard_factor})):
\bea
C\left(\alpha_S\right) &\equiv& 1 \, + \,
C_F \sum_{n=1}^\infty
\left( \frac{\alpha_S}{\pi} \right)^n C^{(n)}
\, =
\nonumber\\
&=& 1 \, + \, \frac{C_F \, \alpha_S}{\pi} \, C^{(1)}
\, + \, C_F
\left( \frac{\alpha_S}{\pi} \right)^2 C^{(2)}
\, + \, \cdots.
\eea
Since%
%%%%%%%%%%
\footnote{
Furthermore, $\Phi_M$ is not expected
to contain large infrared logarithms
in its perturbative expansion,
$\left| \log(\chi_M) \right| \lsim 1$,
as $\chi_M = \mathcal{O}(1)$.
}:
%%%%%%%%%%
\beq
\Phi_M \, = \, 1 \, + \,
\mathcal{O}\left(\alpha_S\right),
\eeq
the first-order coefficient explicitly reads:
\beq
\label{eq_unitary_C_Rem}
C^{(1)} \, = \,
- \, \mathrm{Rem}^{(1)}_M
\, = \, - \,
\mathrm{Rem}^{(1)}\left( \chi=\chi_M \right).
\eeq
The following remarks are in order:
%%%%%%%%%%%%%%%%%
\begin{enumerate}
%%%%%
\item
The first-order coefficient function just equals minus the partially-integrated Remainder function
evaluated at the maximal correlation
angle considered, namely
$\chi=\chi_M$.
That implies in particular that,
in the massive case, $C^{(1)}$ does depend,
in addition to the mass parameter $\eta$,
also on $\chi_M$:
\beq
C^{(1)} \, = \, C^{(1)}\left( \chi_M; \eta \right).
\eeq
%%%%%
\item
The first-order ``final'' Remainder function is equal to the previous ones:
\beq
\mathrm{Rem}^{(1)}\left( \chi \right)
\, = \,
\mathrm{Rel}^{(1)}\left( \chi \right)
\, = \,
\mathrm{Res}^{(1)}\left( \chi \right).
\eeq
%%%%%
\item
Eq.(\ref{eq_EEC_EF_main_eq}) exactly satisfies the unitarity
equation (\ref{eq_new_EF_unitarity}).
%%%%%%%%%%%%%%%
\end{enumerate}

%%%%%%%%%%%%%%%%%%%%%%%%%%%%%%%%%%%%
{\color{black}
\section{Massive EEC to first order}
\label{sec_EEC_FO}

In this section we present an explicit 
calculation of the EEC correlation in
the massive case to $\mathcal{O}(\alpha_S)$.
This involves both quark-gluon and quark-antiquark correlations,
which are treated in the next two sections.
In sec.$\,$\ref{sec_Compare} 
we compare our results with previous results
in the literature.

%%%%%%%%%%%%%%%%%%%%%%%%%%%%%%%%%

\subsection{Quark-gluon correlation}
\label{sec_Qg}
}
Because of the charge-conjugation symmetry of QCD,
the antiquark-gluon EEC correlation is equal to the quark-gluon one.
Therefore, to get the total, we simply have to evaluate the $Qg$
correlation only, and then multiply the result by a factor two.

The first step is to express the quark-gluon angle
as a function of the quark and antiquark
energies {\color{black}(defined in 
eqs.(\ref{eqs_def_energ}) in
Appendix$\,$\ref{sec_kino})}:
\beq
\cos\left(\overline{\theta}_{Qg} \right) \, = \,
f\left( x, \overline{x}; \, \eta\right);
\eeq
where:
\beq
\label{def_eq_funz_f}
f\left( x,\overline{x}; \, \eta \right)
\, \equiv \,
\frac{2 - 2 x - 2 \overline{x} + x \, \overline{x} + x^2}{
\left(2 - x - \overline{x} \right) \sqrt{x^2 \, - \, \eta^2} }.
\eeq
We have defined:
\bea
\overline{\theta}_{Q g} &\equiv&
\pi \, - \, \theta_{Qg}.
\eea
The above-defined angle vanishes in the
back-to-back or 2-jet limit
($\chi \mapsto 0^+$).
The EEC kinematical constraint then reads:
\beq
\delta\big[
\cos(\chi)
\, - \, f\left( x, \overline{x}; \, \eta \right)
\big];
\eeq
Since $\overline{x}$ (unlike $x$) does not appear
inside any radical
on the r.h.s. of eq.(\ref{def_eq_funz_f}),
the integration
over $\overline{x}$ turns out to be much simpler
than in the quark-antiquark correlation
case (see next section).
In the present case, there is indeed a unique solution of the equation
$f\left( x,\overline{x}; \eta\right)
= \cos(\chi)$ in $\overline{x}$, which reads:
\bea
\label{eq_star_below}
\overline{x}_*  \, = \,
\overline{x}_*(x, \cos\chi; \,\eta)  \, = \,
2 \, - \, x \, - \, \frac{2(1\, - \, x)}{
2\, -\, x\, - \, \sqrt{x^2 \, - \, \eta^2} \,\cos\chi}.
\eea

%%%%%%%%%%%%%%%%%%%%%%%%%%%%%%%%%%%%%%%%%%%

\subsubsection{\texorpdfstring{$Qg$}{Qg} differential distribution}

By comparing the last member of the above equation
with the r.h.s. of eqs.(\ref{eqs_xmin_and_xmax}), it is immediate
to check that it always holds:
\beq
\overline{x}_{\min} \, \le \, \overline{x}_* \, \le \, \overline{x}_{\max},
\eeq
so there is no restriction
on the range of the (external) $x$-integration
($x \in [\eta,1]$),
{\color{black}related to the action of the Dirac $\delta$-function
($\overline{x}_{\min}$ and $\overline{x}_{\max}$
are defined in eqs.(\ref{eqs_xmin_and_xmax})
in Appendix$\,$\ref{sec_kino}).}
The integration of the $\delta$-function
produces the Jacobian factor
(which turns out to be rather singular in
the numerical calculation):
\beq
\left(
\left| \frac{\partial f}{\partial \overline{x}}\right|_{\overline{x}\mapsto \overline{x}_*(x)}
\right)^{-1}
\, = \,
\frac{ 2(1 \, - \, x) \, \sqrt{x^2 \, - \, \eta^2}}{
\left(2 \, - \, x \, - \, \sqrt{x^2 \, - \, \eta^2} \cos\chi
\right)^2
}.
\eeq
Note that:
\beq
\frac{E \, \omega}{Q^2} \, = \, 
\frac{x(2-x-\overline{x})}{4}.
\eeq
We have then to include
a factor two coming from the consideration
of the two distinct $(Q,g)$
and $(g,Q)$ pairs, and another factor two
from the inclusion of the $\overline{Q}g$ correlation.

Therefore the quark-gluon + antiquark-gluon contributions
to the differential EEC function
in the massive case finally read:
\bea
\label{Qg_eval_to_num}
\frac{1}{\sigma_{CC}^{(0)}}
\frac{d\Sigma^{(CC)}_{Qg+\overline{Q}g}}{d\cos\chi}
&=&
\frac{C_F \alpha_S}{2\pi}
\int\limits_\eta^1 dx
\int\limits_{\overline{x}_{\min}(x)}^{\overline{x}_{\max}(x)}
d \overline{x}
\, \delta\left[
\cos(\chi)
- f\left( x,\overline{x}; \, \eta\right)
\right]
x \, \left( 2 - x - \overline{x} \right) \,
\mathcal{M}_{CC}\left(x,\overline{x};\eta \right) \, =
\nonumber\\
&=&
\frac{C_F \alpha_S}{2\pi}
\int\limits_\eta^{1} dx 
\Bigg\{
\left| \frac{\partial f}{\partial \overline{x}}
\right|^{-1}
 x \left( 2 - x - \overline{x} \right)
\mathcal{M}_{CC}\left(x,\overline{x};\eta \right)
\Bigg\}_{\overline{x} \mapsto \overline{x}_*(x)} (C=V,A). \!
\eea
{\color{black}The matrix elements $\mathcal{M}_{CC}$
are given explicitly in Appendix \ref{sec_ME},
see eqs.(\ref{eq_ME_VV}) and (\ref{eq_ME_AA}).}
The above one-dimensional integral,
on a finite domain, is easily evaluated
numerically, basically with arbitrary precision.

%%%%%%%%%%%%%%%%%%%%%%%%%%%%%%%%%%%%%%%%%%%%%%%%%%%

\subsubsection{\texorpdfstring{$Qg$}{Qg} partially-integrated distribution}

By integrating the differential distribution
of the EEC correlation over $\cos\chi$,
one obtains:
\bea
&&\frac{1}{\sigma_{CC}^{(0)}}
\int\limits_{-1}^{\cos\chi}
\frac{d\Sigma^{(CC)}_{Qg+\overline{Q}g}}{d\cos\chi'}
\, d\cos\chi'
\nonumber\\
&=& \frac{C_F \alpha_S}{2\pi}
\int\limits_{\eta}^{1} dx
\int\limits_{\overline{x}_{\min}(x)}^{\overline{x}_{\max}(x)}
d\overline{x} \,
 x \, \left( 2 - x - \overline{x} \right) \,
\mathcal{M}_{CC}\left(x,\overline{x};\, \eta\right) \,
\theta\left[
\cos(\chi)
\, - \, f\left( x,\overline{x}; \, \eta\right)
\right] \, =
\nonumber\\
&=& \frac{C_F \alpha_S}{2\pi} \int\limits_{\eta}^1 dx
\int\limits_{\overline{x}_*}^{\overline{x}_{\max}}
d\overline{x} \,
 x \, \left( 2 - x - \overline{x} \right) \,
\mathcal{M}_{CC}\left(x,\overline{x}; \, \eta\right) \, =
\\ \nonumber
&=& \frac{C_F \alpha_S}{2\pi} \int\limits_0^{1-\eta} dy
\int\limits^{\overline{y}_*}_{\overline{y}_{\max}}
d\overline{y} \,
\left( 1 - y \right) \, \left(y \, + \, \overline{y}\right) \,
\mathcal{M}_{CC}\left(1-y,1-\overline{y};\, \eta\right);
\eea
where $\theta(z)\equiv 1$ for $z>0$
and zero otherwise is the standard Heaviside step
function.
We have defined:
\bea
y &=& 1 \, - \, x;
\nonumber\\
\overline{y} &=& 1 \, - \, \overline{x}.
\eea
The first (inner) integral, over $\overline{x}$
(or $\overline{y}$),
is easily evaluated analytically.
The integrand is naturally written
as a (finite) Laurent series in $\overline{y}$;
for the vector case, for example:
\bea
&& (1-y)\left(y \, + \, \overline{y}\right) \,
\mathcal{M}_{VV}\left(1-y,1-\overline{y};\, \eta\right) \, =
\nonumber\\
&=& \frac{(1-y)\left(y \, + \, \overline{y}\right)}{ \sqrt{1 - \eta^2} }
\Bigg[
\frac{ (1-y)^2 + (1-\overline{y})^2 - \eta^2  }{
y \, \overline{y} }
\, - \, \frac{\eta^2}{2+\eta^2}
\frac{ y^2 + \overline{y}^2 }{y \, \overline{y}}
\, - \, \frac{\eta^2}{2}
\left(
\frac{ 1 }{y^2} \, + \, \frac{ 1 }{\overline{y}^2} \right)
\Bigg] \, =
\nonumber\\
&=& \sum_{n=-2}^2 R_n^{(VV)}(y) \, \overline{y}^n,
\eea
where $R_n^{(VV)}(y)$ are, in general, rational functions of $y$ (i.e. also negative powers of $y$ are
involved).
The integral over $\overline{y}$
is elementary:
\bea
\sum_{n=-2}^2
R_n^{(VV)}(y)
\int\limits^{\overline{y}_*(y)}_{\overline{y}_{\max}(y)}
d\overline{y} \,\overline{y}^n
&=& R_{-2}^{(VV)}(y)
\left[
\frac{1}{\overline{y}_{\max}(y)}
\, - \, \frac{1}{\overline{y}_*(y)}
\right]
\, + \, R_{-1}^{(VV)}(y)
\log\left[
\frac{\overline{y}_*(y)}{\overline{y}_{\max}(y)}
\right]
\, +
\nonumber\\
&+&
\sum_{n=0}^2 \frac{ R_n^{(VV)}(y)}{n+1} \,
\Big\{
\left[ \overline{y}_*(y) \right]^{n+1}
\, - \,
\left[ \overline{y}_{\max}(y) \right]^{n+1}
\Big\}.
\eea
The second (external) integration, over $x$ (or $y$), is easily evaluated numerically.

By fitting the partially-integrated distribution 
(for example with a polynomial)
and then differentiating the fitted function with respect
to $\cos\chi$, we find complete agreement with the direct numerical evaluation of the differential distribution.

%%%%%%%%%%%%%%%%%%%%%%%%%%%%%%%%%%%%%

\subsection{Quark-antiquark correlation}
\label{sec_QQbar}

In order to express the quark-antiquark angle,
$\theta_{Q\overline{Q}}$,
as a function of $x$ and $\overline{x}$,
let us use the equation
of energy conservation
\beq
z \, = \, 2 \, - \, x \, - \, \overline{x}
\eeq
{\color{black} inside eq.(\ref{eq_p_plus_pb})
(see Appendix \ref{sec_kino}),
to obtain}:
\beq
2 \left( x \, + \, \overline{x} \, - \, 1 \right) \, = \, \eta^2 \, + \, x \, \overline{x} \, + \,
\sqrt{\,\left(x^2 \, - \, \eta^2\right) \,
\left(\overline{x}^2 \, - \, \eta^2\right)\,}
\, \cos\overline{\theta}_{Q\overline{Q}};
\eeq
where we have defined:
\beq
\overline{\theta}_{Q\overline{Q}} \,\, \equiv \,\,
\pi \, - \, \theta_{Q\overline{Q}}.
\eeq
In order to have a hint of the kinematic situation,
let us begin by considering the massless limit,
in which the (much simpler) equation is obtained:
\beq
2 \left( x \, + \, \overline{x} \, - \, 1 \right)
\, = \, x \, \overline{x} \,
\left(
1 \, + \, \cos \overline{\theta}_{Q\overline{Q}}
\right)
\qquad (m=0).
\eeq
By solving the above equation with respect to the $Q\overline{Q}$ angle,
one obtains:
\beq
\cos \overline{\theta}_{Q\overline{Q}}
\, = \, \frac{ 2 \left( x \, + \, \overline{x} \, - \, 1 \right)
\, - \, x \, \overline{x}}{x \, \overline{x}}
\qquad (m=0).
\eeq
In the massive case, the formula for the quark-antiquark angle
in terms of the quark and antiquark energies reads:
\beq
\label{eq_costhqgxxbar}
\cos \overline{\theta}_{Q\overline{Q}}
\, = \, g\left(x,\overline{x};
\, \eta \right);
\eeq
where:
\beq
g\left( x, \overline{x}; \, \eta \right)
\, \equiv \,
\frac{ 2 \left(x \, + \, \overline{x} \, - \, 1\right) \,- \, x \, \overline{x}
\, - \, \eta^2 }{
\sqrt{\left(x^2 \, - \, \eta^2\right) \left(\overline{x}^2 \, - \, \eta^2\right) }}.
\eeq
Note the symmetry of the r.h.s. of the
above equation under exchange of the
quark and antiquark energies
($x \leftrightarrow \overline{x}$).
The EEC kinematic constraint
on the quark-antiquark angle,
\beq
\delta\left( \cos \overline{\theta}_{Q\overline{Q}}
\, -\, \cos\chi \right),
\eeq
is then explicitly written in terms of $x$
and $\overline{x}$:
\beq
\label{eq_delta_F}
\delta\big[
\cos\chi \, - \,
g\left(x,\overline{x};
\, \eta \right)
\big].
\eeq
Let us remark that,
since we do intend to integrate over $\overline{x}$
first, the function $g$
is to be thought as an explicit function of
$\overline{x}$
(the quantities $x,\cos\chi$ and $\eta$
have then the role of parameters upon which $g$ depends).

\noindent
As we have found above, in the massless limit, the function $g$
largely simplifies into:
\beq
\label{eq_F_massless}
g_0\left( x, \overline{x} \right)
\, \equiv \,
g\left( x, \overline{x}; \, \eta = 0 \right)
\, \equiv \,
\frac{ 2 x \, + \, 2 \overline{x} \, - \, 2 \,- \, x \, \overline{x}}{
x \overline{x}}
\qquad (m=0).
\eeq
In the massless case, there is a {\it unique} solution in $\overline{x}$,
which is provided by:
\beq
\label{eq_xbarstar_massless}
\left( \overline{x}^* \right)_0 \, = \,
\frac{2(1 \, - \, x)}{2 \, - \,
\left( 1 \, + \, \cos\chi \right) x }
\qquad (m=0).
\eeq
It is easy to check that, in the massless case,
for any allowed values of $x$ and $\chi$,
it holds:
\beq
\left( \overline{x}_{\min} \right)_0 \, = \, 1 \, - \, x
\,\, \le \,\, \left( \overline{x}^* \right)_0
\,\, \le \,\, \left( \overline{x}_{\max} \right)_0 \, = \, 1
\qquad (m=0).
\eeq
Therefore, in the massless case, the integration
over $\overline{x}$ of the $\delta$-function EEC kinematic
constraint does not lead to any restriction
on the values of $x \in [0,1]$.

In order to solve the ``massive'' equation
in $\overline{x}$
implied by the $\delta$-function constraint (\ref{eq_delta_F}), namely:
\beq
\label{eq_qqbarcorr_true}
\frac{ 2 x + 2 \overline{x} - 2 - x \, \overline{x}
- \eta^2 }{
\sqrt{\left(x^2 - \eta^2\right) \left(\overline{x}^2 - \eta^2\right) }}
\, = \, \cos\chi,
\eeq
one has first to eliminate the square root
involving $\overline{x}$.
Therefore one has to square
at both sides of the above equation, introducing, in addition
to the ``good'', i.e. true, solution, a spurious solution,
which has later to be discarded ---
namely the solution of the ``spurious'' equation
\beq
\frac{ 2 x + 2 \overline{x} - 2 - x \, \overline{x}
- \eta^2 }{
\sqrt{\left(x^2 - \eta^2\right) \left(\overline{x}^2 - \eta^2\right) }}
\, = \, - \, \cos\chi
\qquad (\mathrm{spurious\,\,equation}).
\eeq
We then obtain the second-order equation
in $\overline{x}$:
\beq
\label{eq_squared}
\left( \overline{x}^2 \, - \, \eta^2 \right) \, \xi^2
\, = \,
\left(
- \, 2 \, - \, \eta^2 \, + \, 2 x \, + \, 2 \overline{x} \, - \, x \, \overline{x}
\right)^2 ;
\eeq
where we have defined the (frequently-occurring) quantity:
\beq
\xi \, = \, \xi\left( x, \cos\chi; \, \eta \right)
\, \equiv \,
\sqrt{ x^2 \, - \, \eta^2 } \, \cos\chi.
\eeq
The solutions of eq.(\ref{eq_squared}) are given by
(note that the asterisks are superscripts in the present case):
\bea
\label{eq:true_sol}
\overline{x}^* &=& \overline{x}^*\left(x,\chi;\, \eta\right)
\, =
\\ \nonumber
&=& \frac{ (2 \, - \, x) \big[ 2 (1 \, - \, x) \, + \, \eta^2 \big]
 \, + \, \cos\chi \,
\sqrt{ x^2 \, - \, \eta^2 } \, \sqrt{ 4 (1 \, - \, x)^2 \, - \, \eta^2
\left( x^2 \, - \, \eta^2 \right) \sin^2\chi  }
}{ (2 \, - \, x)^2 \, - \,
\left( x^2 \, - \, \eta^2 \right)
\, \cos^2\chi };
\qquad
\eea
and:
\bea
\label{eq:also_true_sol}
\overline{x}^{**} &=& \overline{x}^{**}\left(x,\chi;\, \eta\right) \, =
\\ \nonumber
&=& \frac{ (2 \, - \, x) \big[ 2 (1 \, - \, x) \, + \, \eta^2 \big]
 \, - \, \cos\chi \, \sqrt{ x^2 \, - \, \eta^2 } \,  \sqrt{ 4 (1 \, - \, x)^2 \, - \, \eta^2
\left( x^2 \, - \, \eta^2 \right) \sin^2\chi  }
}{ (2 \, - \, x)^2 \, - \,
\left( x^2 \, - \, \eta^2 \right)
\, \cos^2\chi }.
\qquad
\eea
The following remarks are in order:
%%%%%%%%%%%%%%%%%
\begin{enumerate}
%%%%%
\item
The first solution above, namely $\overline{x}^*$,
is the one having the (ab-initio) massless limit,
eq.(\ref{eq_xbarstar_massless}), as:
\beq
\lim_{\eta\to 0^+} \overline{x}^*\left(x,\chi;\, \eta\right)
\, = \, \frac{ 2(1-x)(2 - x + x \cos\chi)
}{(2-x)^2 - x^2 \cos^2\chi}
\, = \, \frac{2(1-x)}{2 \, - \,
\left( 1 \, + \, \cos\chi \right) x}.
\eeq
We then expect the first solution to be,
in some sense, the ``dominant'' one for
small masses, namely $\eta \ll 1$.
As we are going to show later, this expectation
turns out to be correct {\it a posteriori}.
%%%%%
\item
\label{item_compl_dom}
A {\it crucial point} is that
we do not know {\it a priori} in which regions
of the $(\chi,x)$ plane, $\overline{x}^*$ and
$\overline{x}^{**}$ are actual solutions
of the original EEC kinematic-constraint
equation (\ref{eq_qqbarcorr_true});
By solving the (weaker) squared equation (\ref{eq_squared}),
instead of the original one
(\ref{eq_qqbarcorr_true}),
such information is indeed lost.
%%%%%%%%%%%%%%%%% inizio enumerazione interna
\begin{enumerate}
%%%%%
\item
Let us consider $\overline{x}^*$ first.
We substitute its expression,
namely the r.h.s. of eq.(\ref{eq:true_sol}), into the
original equation (\ref{eq_qqbarcorr_true})
and evaluate numerically the resulting expression
on a fine grid (a large number of points)
in the $(\chi,x)$
plane, with $0\le \chi \le \pi$
and $\eta \le x \le 1$.
We find that $\overline{x}^*$
{\it is not} a (generally complex) {\it solution}
of eq.(\ref{eq_qqbarcorr_true})
in the (small) {\it top right} rectangular region
(see fig.\,\ref{fig_plotsolregxbstar})

\beq
\label{diseq_constraint_addiz}
\tilde{x}(\eta) \, \le \, x \, \le \, 1;
\quad
\frac{\pi}{2} \, \le \, \chi \, \le \, \pi:
\quad
\overline{x}^* \,\, \mathrm{no\,\,solution}
\eeq
The function $\tilde{x}(\eta)$,
defined in eq.(\ref{eq_-def_xtilde}), has a minimum at:
\beq
\eta_{\min} \, = \, 2 - \sqrt{2} \, = \, 0.585786\cdots;
\eeq
where it takes the value
\beq
\tilde{x}(\eta=\eta_{\min}) \, = \, 2
\left( \sqrt{2} \, - \, 1 \right) \, = \, 0.828427 \cdots.
\eeq
The ``kinematic origin'' of the ``forbidden region''
(\ref{diseq_constraint_addiz})
above, is not hard to understand:
if the $Q\overline{Q}$ relative angle is small,
then the gluon
basically recoils against a quasi-collinear
quark-antiquark pair.
In this situation,
the maximal quark energy $x$ corresponds
to the case in which the antiquark
is soft, $\overline{x} \gsim \eta$,
i.e. its space momentum is small,
$\left| \vec{p}_{\overline{Q}} \right| \ll m$.
Therefore one has basically
an antiquark at rest,
or almost at rest,
and a quark recoiling against a gluon.
The total energy of the latter quark-gluon pair
is not $Q$, but rather $\lsim Q-m$, hence $x \sim \tilde{x}(\eta)$.
In particular, in this configuration, the quark energy $x$
cannot reach one, as it happens instead
when the $Q\overline{Q}$ angle is large,
i.e. when the quark and the
antiquark are basically back-to-back
and the gluon is soft.
%%%%%

%
%%%%%%%%%%%%%%%%%%%
%
\begin{figure}[h]
\begin{center}
\includegraphics[width=0.9\textwidth]{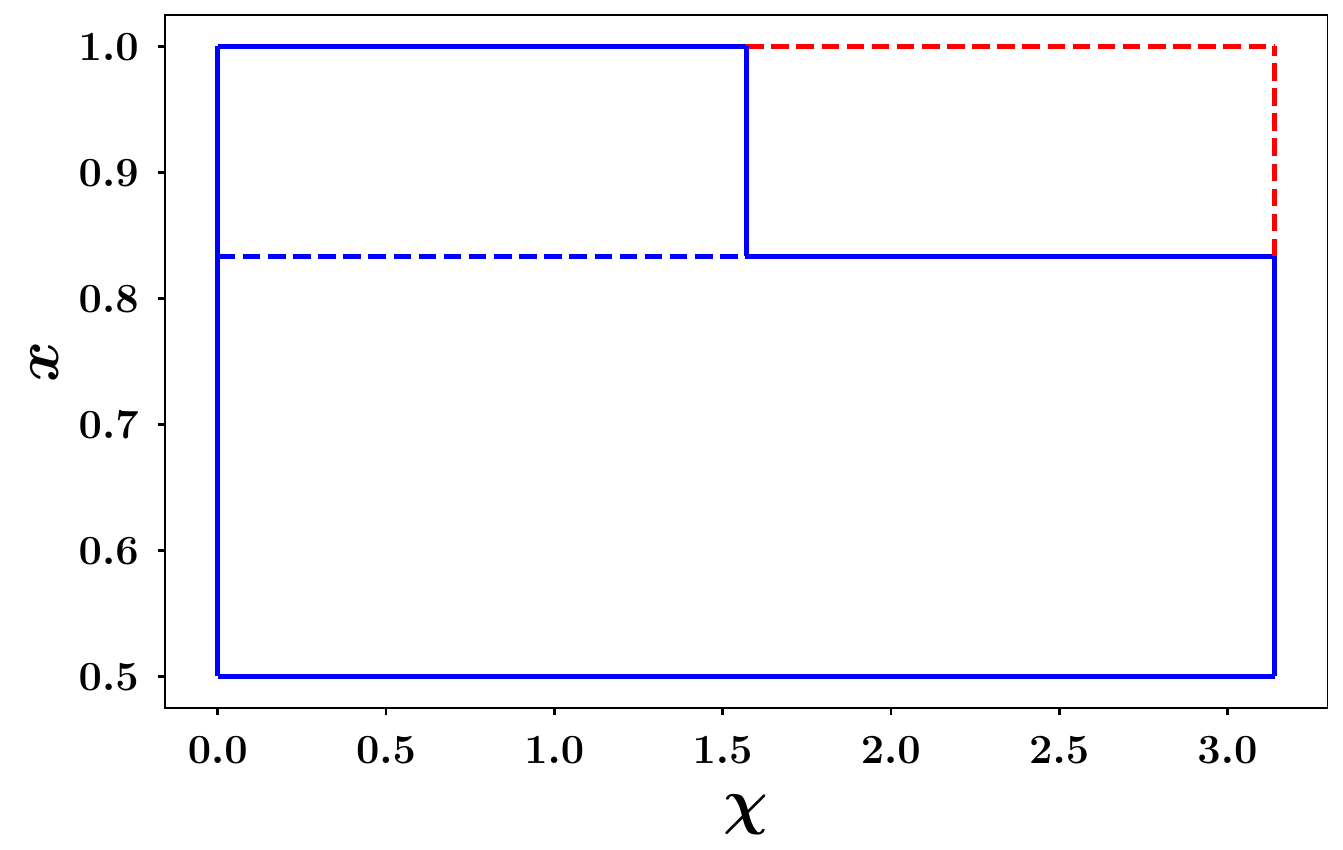}
\footnotesize
\caption{
\label{fig_plotsolregxbstar}
\it
The region inside the continuous blue contour
is the region, in the plane $(\chi,x)$,
where $\overline{x}^*$, eq.(\ref{eq:true_sol}), is an actual solution
of the (original) kinematic-constraint
equation (\ref{eq_qqbarcorr_true}),
for $\eta=0.5$
(in other words, $\overline{x}^*$
is not a solution of eq.(\ref{eq_qqbarcorr_true})
inside the (top right) small rectangle with the red
dashed sides).
$\overline{x}^{**}$, eq.(\ref{eq:also_true_sol}), is a solution of eq.(\ref{eq_qqbarcorr_true}) only
in the (top left) small rectangle above the
blue dashed side.
}
\end{center}
\end{figure}
%
%%%%%%%%%%%
%
%%%%%
\item
Let us now consider the {\it second solution}
$\overline{x}^{**}(x,\chi;\, \eta)$, given in eq.(\ref{eq:also_true_sol}).
By using the same method above,
we find that $\overline{x}^{**}$
{\it is a solution} of the original equation
(\ref{eq_qqbarcorr_true})
in the small {\it top left} rectangle
in the $(\chi,x)$ plane
(see fig.\,\ref{fig_plotsolregxbstar})
\beq
\label{diseq_constraint_addiz2}
0 \, \le \, \chi \, \le \, \frac{\pi}{2};
\quad
\tilde{x}(\eta)
\, \le \, x \, \le \, 1:
\quad
\overline{x}^{**} \,\, \mathrm{is \,\, solution}.
\eeq
%%%%%%%%%%%%%%% fine enumerazione interna
\end{enumerate}
%%%%%
\item
By comparing the regions where
$\overline{x}^*$ and $\overline{x}^{**}$
are solutions of the original equation,
we can make the following ``geometrical''
observations.

The ``large'' rectangle
\beq
\label{large_rect}
0 \, \le \, \chi \, \le \, \pi;
\qquad
\eta \, \le \, x \, \le \, \tilde{x}(\eta)
\eeq
is somehow associated to the massless limit.
Indeed, in the massless limit,
\beq
\lim_{\eta \to 0^+} \tilde{x}(\eta) \, = \, 1,
\eeq
and this region invades the whole space,
\beq
0 \, \le \, \chi \, \le \, \pi;
\qquad
0 \, \le \, x \, \le \, 1.
\eeq
Furthermore, in the rectangle (\ref{large_rect}),
only $\overline{x}^*$ is a solution of the original
equation.

Let us now consider the upper horizontal
rectangle in the
$(\chi,x)$ plane:
\beq
\label{strip_to_vanish}
0 \, \le \, \chi \, \le \, \pi;
\qquad
\tilde{x}(\eta) \, \le \, x \, \le \, 1.
\eeq
Both $\overline{x}^*$ and $\overline{x}^{**}$
are solutions in the {\it left half} of this
strip, namely the region
\beq
\label{eq_small_rect}
0 \, \le \, \chi \, \le \, \frac{\pi}{2};
\quad
\tilde{x}(\eta) \, \le \, x \, \le \, 1:
\quad
\overline{x}^* \,\, \mathrm{and} \,\, \overline{x}^{**}
\,\, \mathrm{solutions}.
\eeq
Furthermore, the rectangle
(\ref{eq_small_rect}) is the
only region, in the $(\chi,x)$ plane,
having {\it two} different solutions.

On the contrary, neither $\overline{x}^*$ nor $\overline{x}^{**}$
is a solution of the original equation in the right half of the strip, namely the region
\beq
\frac{\pi}{2} \, \le \, \chi \, \le \, \pi;
\quad
\tilde{x}(\eta) \, \le \, x \, \le \, 1:
\quad
\overline{x}^* \,\, \mathrm{and} \,\, \overline{x}^{**}
\,\, not \mathrm{\,\, solutions}.
\eeq
It looks {\it as if} the long strip (\ref{strip_to_vanish}) is folded  around
its middle vertical segment $\chi=\pi/2$,
with the right part going over the left one ---
that is a rigid rotation of the right part by $180^\circ$.
%%%%%

%
%%%%%%%%%%%%%%%%%%%
%
\begin{figure}[h]
\begin{center}
\includegraphics[width=0.9\textwidth]{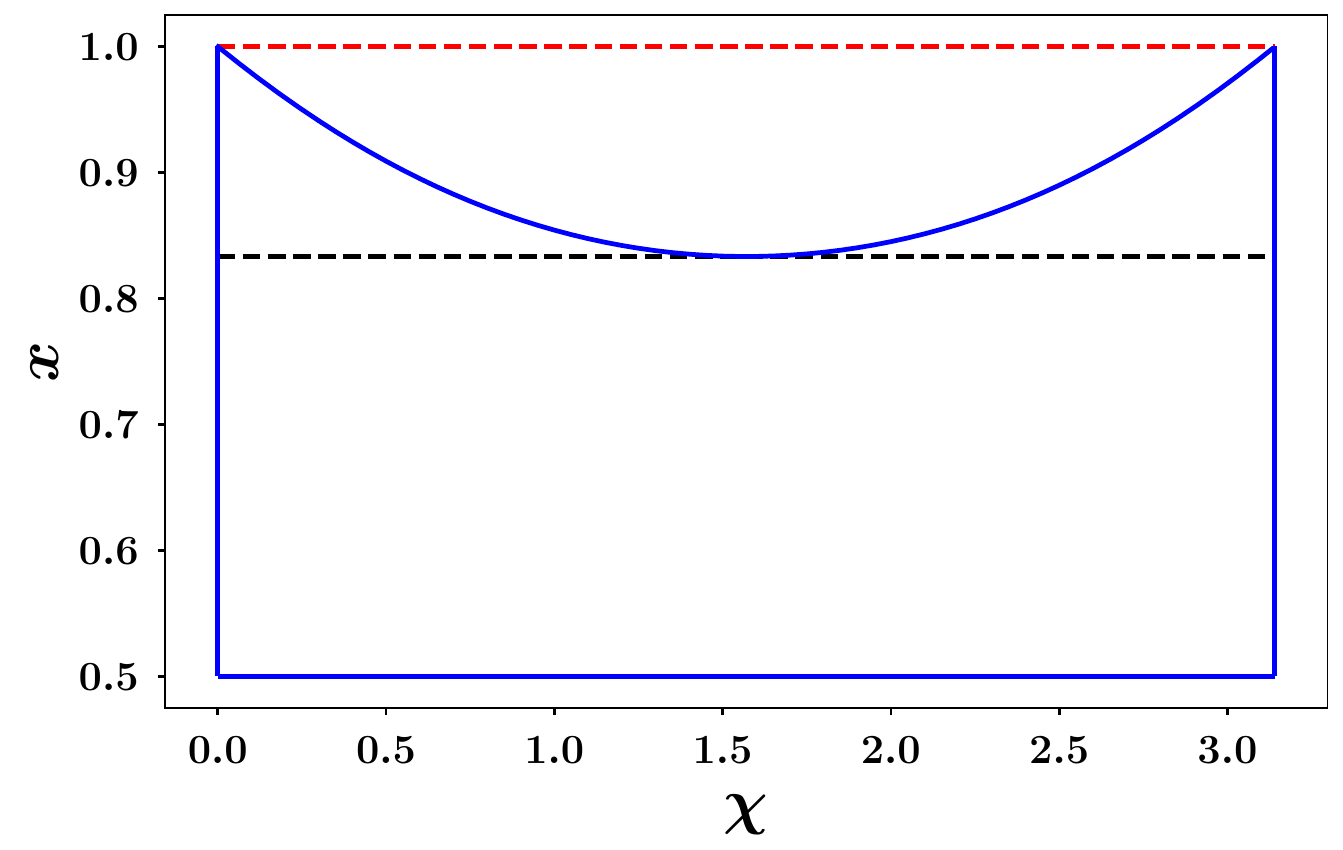}
\footnotesize
\caption{
\label{fig_plotradreal}
\it
The $\overline{Q}$ energies $\overline{x}^*$ and $\overline{x}^{**}$
are real only below the blue curve ($\eta=0.5$).
Note that the horizontal black dashed line
$x=\tilde{x}(\eta)$ is tangent
to the blue curve at its minimum
at $\chi=\pi/2$.
}
\end{center}
\end{figure}
%
%%%%%%%%%%%
%

%%%%%
\item
\label{item_real_energ}
The formulae of $\overline{x}^*$ and $\overline{x}^{**}$,
eq.(\ref{eq:true_sol}) and (\ref{eq:also_true_sol}) respectively,
involve the square root of an expression,
namely
\beq
\label{eq_expression}
(1 \, - \, x)^2 \, - \, \frac{\eta^2}{4} \left(x^2 \, - \, \eta^2\right) \, \sin^2\chi,
\eeq
which can become negative
(for example at $x=1$)%
%%%%%%%%%%
\footnote{
This problem disappears in the massless limit, as
\beq
\nonumber
(1 \, - \, x)^2 \, - \, \frac{\eta^2}{4} \left(x^2 \, - \, \eta^2\right) \, \sin^2\chi
\,\, \mapsto \,\,
(1 \, - \, x)^2,
\eeq
in which the argument of the radical becomes
positive-definite, as $(1-x)^2 \ge 0$.
}.
%%%%%%%%%
Since $\overline{x}^*$ and $\overline{x}^{**}$
are (antiquark) energies, they have to be real,
implying that the argument of the radical
above, the expression (\ref{eq_expression}), has to be positive.
That turns out to provide a non-trivial upper bound
on the quark energy $x$%
%%%%%%%%%%
\footnote{
The antiquark energy in the simpler quark-gluon
correlation $\overline{x}_*$
(eq.(\ref{eq_star_below}) in the previous section)
is instead always real.
}
%%%%%%%%%%
(see fig.\ref{fig_plotradreal}):
\bea
\label{eq_expression_x_max}
\eta \,\, \le \,\, x \,\, \le \,\, \hat{x}_{\max} ;
\eea
where:
\beq
\hat{x}_{\max} \, = \,
\hat{x}_{\max}\left( \chi; \, \eta \right)
\, \equiv \,
\frac{1 \, - \, \eta/2 \, \sin\chi \,
\sqrt{1 \, - \, \eta^2 \, + \, \eta^4/4  \, \sin^2\chi   }
}{1 \, - \, \eta^2/4 \, \sin^2\chi}.
\eeq
Note that we could also write:
\beq
\hat{x}_{\max} \, = \, \hat{x}_{\max}\left(\sin\chi;
\, \eta \right).
\eeq
$\hat{x}_{\max}\left(\chi;\, \eta\right)$
reaches its (absolute) minimum in the
midpoint $\chi=\pi/2$:
\beq
\hat{x}_{\max}\left(\chi=\frac{\pi}{2};
\, \eta \right)
\, = \, \tilde{x}(\eta)
\,\, \le \,\,
\hat{x}_{\max}\left(\chi; \, \eta\right),
\qquad \forall \chi \, \in \, [0,\pi].
\eeq
It is easy to show analytically that:
\beq
\eta \, \le \, \tilde{x}(\eta).
\eeq
By combining the two above inequalities,
it follows that it always holds:
\beq
\eta \, \le \,
\hat{x}_{\max}\left(\chi ; \, \eta\right).
\eeq
%%%%%
\item
In previous item \ref{item_compl_dom}, we have determined
in which regions $\overline{x}^*$ and $\overline{x}^{**}$
are general complex solution of the kinematic-constraint
equation (\ref{eq_qqbarcorr_true}).
In order to determine their physical (real) domains,
we have to subtract from the above regions
the sub-regions where $\overline{x}^*$ and $\overline{x}^{**}$
are not real, as determined in item \ref{item_real_energ}.
%
%%%%%%%%%%%%%%%%% inizio lista interna
\begin{enumerate}
%%%%%
\item
In the case of $\overline{x}^*$,
we obtain the following physical domain
$D^*$
(see fig.\,\ref{fig_plotradrealimpr}):
\beq
0 \le \chi \le \pi;
\qquad
\eta \, \le \, x \, \le \, x_{\max}\left( \chi; \, \eta \right);
\eeq
 where:
\beq
\label{eq_expression_x_max_tot}
x_{\max}\left( \chi; \, \eta \right)
\, \equiv \,
\theta\left( \frac{\pi}{2} \, - \, \chi \right) \,
\hat{x}_{\max}\left(\chi ; \, \eta\right)
\, + \,
\theta\left( \chi \, - \, \frac{\pi}{2} \right)
\tilde{x}(\eta) .
\eeq
Note that the above function
$x_{\max}\left( \chi; \, \eta \right)$
is a continuous function of $\chi$,
together with its first derivative,
across the midpoint (junction point) $\chi=\pi/2$.
%%%%%

%
%%%%%%%%%%%%%%%%%%%
%
\begin{figure}[ht]
\begin{center}
\includegraphics[width=0.9\textwidth]{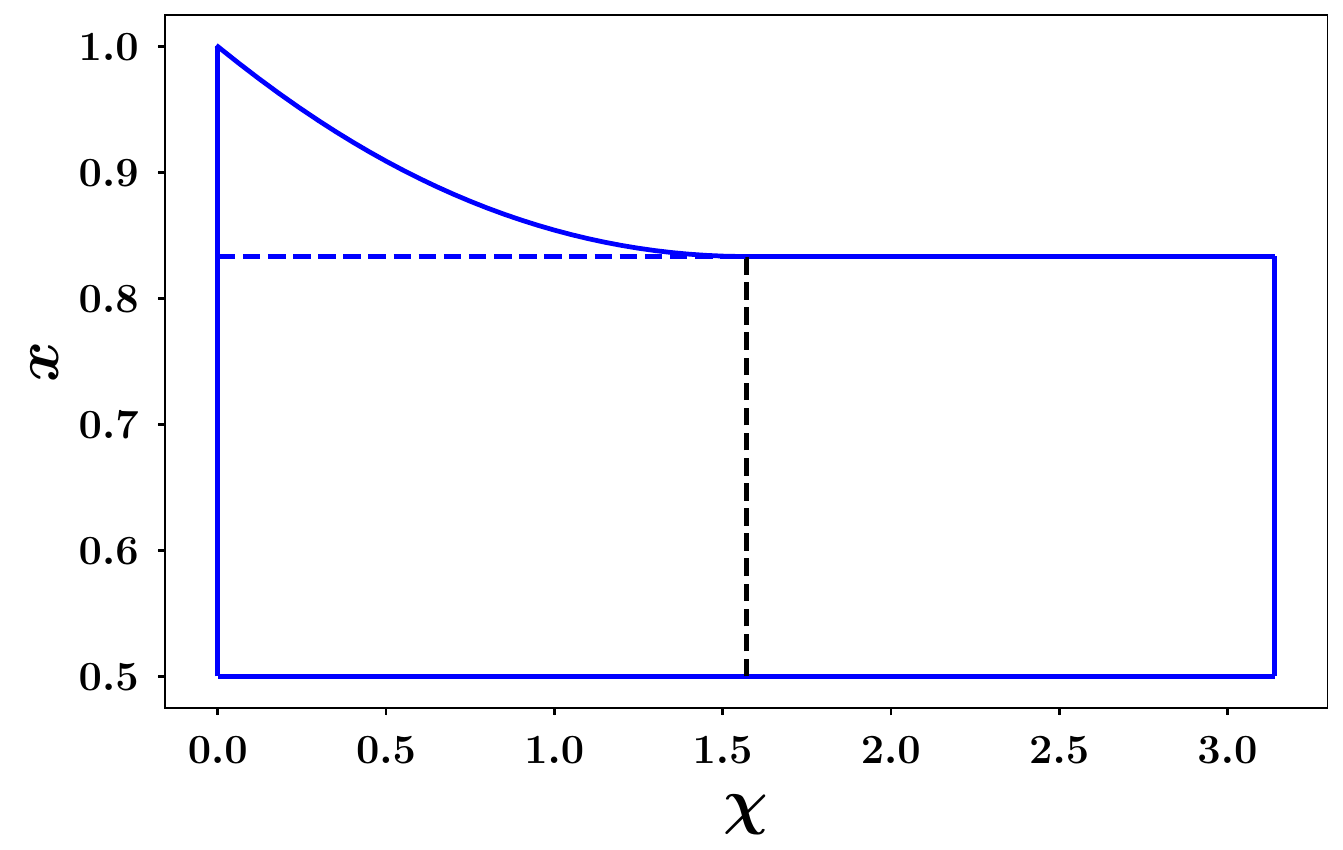}
\footnotesize
\caption{
\label{fig_plotradrealimpr}
\it
The physical (real) domain of $\overline{x}^*$,
for $\eta=0.5$,
is the region contained inside the continuous blue contour.
The physical domain of $\overline{x}^{**}$
is the (small) curvilinear triangle above
the blue dashed line (having the shape
of a main-sail).
}
\end{center}
\end{figure}
%
%%%%%%%%%%%
%

%%%%%
\item
In the case of $\overline{x}^{**}$,
the physical domain $D^{**}$
is the following small curvilinear triangle
--- of main-sail-like shape
(see fig.\,\ref{fig_plotradrealimpr}):
\beq
0 \, \le \, \chi \, \le \, \frac{\pi}{2};
\qquad
\tilde{x}(\eta) \, \le \, x \, \le \,
\hat{x}_{\max}\left( \chi; \, \eta \right).
\eeq
%%%%%%%%%%%%%%% fine lista interna
\end{enumerate}
%%%%%
\item
As far as $\overline{x}^*$ is concerned,
in order for the $\delta$-function to act,
it has to satisfy
the following condition in its physical domain $D^*$:
\beq
\overline{x}_{\min}\left(x;\, \eta\right)
\,\, \le \,\,
\overline{x}^*\left(x,\chi;\, \eta\right)
\,\, \le \,\,
\overline{x}_{\max}\left(x;\, \eta\right).
\eeq
By means of a numerical computation,
we found that the above condition
is always satisfied, for any value of $\chi$
and $\eta$.

As far as $\overline{x}^{**}$ is concerned,
a similar condition has to hold in
its domain $D^{**}$,
\beq
\overline{x}_{\min}\left(x;\, \eta\right)
\,\, \le \,\,
\overline{x}^{**}\left(x,\chi;\, \eta\right)
\,\, \le \,\,
\overline{x}_{\max}\left(x;\, \eta\right).
\eeq
Also in this case, we found that the above
condition is always satisfied.
%
%%%%%%%%%%%%%%%
\end{enumerate}

%%%%%%%%%%%%%%%%%%%%%%%%%%%%%%%%%%%%%%%%%%%%%%%%%

\subsubsection{$Q\overline{Q}$ differential distribution}

Finally, the quark-antiquark contribution
to the differential EEC function in the massive
case reads:
\bea
\label{eq_to_eval_numer}
\frac{1}{\sigma_{CC}^{(0)}}
\frac{d\Sigma^{(CC)}_{Q\overline{Q}}}{d\cos\chi}
&=&
\frac{C_F \alpha_S}{2\pi}
\int\limits_\eta^1 dx
\int\limits_{\overline{x}_{\min}(x)}^{\overline{x}_{\max}(x)}
d \overline{x}
\, \delta\left[ \cos\chi-g\left(\overline{x}\right) \right]
\frac{x \, \overline{x}}{2} \,
\mathcal{M}_{CC}\left(x,\overline{x};\eta \right) \, =
\nonumber\\
&=&
\frac{C_F \alpha_S}{2\pi}
\int\limits_\eta^{x_{\max}} dx \,
\Bigg\{
\left| \frac{\partial g}{\partial \overline{x}}
\right|^{-1} \frac{x \, \overline{x}}{2}
\, \mathcal{M}_{CC}\left(x,\overline{x};\eta \right)
\Bigg\}_{\overline{x} \mapsto \overline{x}^*(x)} \, +
\\ \nonumber
&+&
\frac{C_F \alpha_S}{2\pi} \,
\theta\left(\frac{\pi}{2} - \chi\right)
\int\limits_{\tilde{x}(\eta)}^{\hat{x}_{\max}(\chi,\eta)} dx \,
\Bigg\{
\left| \frac{\partial g}{\partial \overline{x}}
\right|^{-1} \frac{x \, \overline{x}}{2}
\, \mathcal{M}_{CC}\left(x,\overline{x};\eta \right)
\Bigg\}_{\overline{x} \mapsto \overline{x}^{**}(x)};
\eea
where:
\beq
\left| \frac{\partial g}{\partial \overline{x}}
\right|^{-1} \, = \,
\frac{ \left( x^2 \, - \, \eta^2 \right)^{1/2}
\left( \overline{x}^2 \, - \, \eta^2 \right)^{3/2}
}{ \left|
2 \overline{x} (1 \, - \, x) \, - \, \eta^2
\left(2\,-\,x\,-\,\overline{x}\right)
\right|}.
\eeq
For ease of reference:
%%%%%%%%%%%%%%%
\begin{itemize}
%%%%%
\item
$\overline{x}_{\min} \, = \, \overline{x}_{\min}(x; \eta)$
and $\overline{x}_{\max} \, = \, \overline{x}_{\max}(x;\eta)$
are given in eqs.(\ref{eqs_xmin_and_xmax});
%%%%%
\item
$\overline{x}^* \, = \, \overline{x}^*\left(x,\chi; \, \eta\right)$
is given in eq.(\ref{eq:true_sol});
%%%%%
\item
$x_{\max} \, = \,
x_{\max}\left( \chi; \eta \right)$ is given in
eq.(\ref{eq_expression_x_max_tot});
%%%%%
\item
$\hat{x}_{\max}$
is provided by eq.(\ref{eq_expression_x_max}).
%%%%%
\item
$\tilde{x}(\eta)$ is given in eq.(\ref{eq_-def_xtilde});
%%%%%
\item
$\overline{x}^{**}$
is given in eq.(\ref{eq:also_true_sol}).
%%%%%%%%%%%%%
\end{itemize}
The above one-dimensional integrals,
eq.(\ref{eq_to_eval_numer}),
as in the quark-gluon case,
are also easily evaluated numerically.

%%%%%%%%%%%%%%%%%%%%%%%%%%%%%%%%%%%%%%%%%%%%%%%%%%%%%%%%%

\subsubsection{$Q\overline{Q}$ partially-integrated distribution}

By integrating the differential distribution
of the EEC correlation over $\cos\chi$,
one obtains for the partially-integrated
distribution:
\bea
&&\int\limits_{-1}^{\cos\chi}
\frac{1}{\sigma_{CC}^{(0)}}
\frac{d\Sigma^{(CC)}_{Q\overline{Q}}}{d\cos\chi'}
\, d\cos\chi' \, =
\nonumber\\
&=& \frac{C_F \alpha_S}{2\pi}
\int\limits_{\eta}^1 dx
\int\limits_{\overline{x}_{\min}(x)}^{\overline{x}_{\max}(x)}
d\overline{x} \, \frac{x \, \overline{x}}{2} \,
\mathcal{M}_{CC}\left(x,\overline{x};\, \eta\right) \,
\theta\left[\cos\chi-g\left(x,\overline{x};\,
\eta\right)\right] \, =
\\\nonumber
&=& \frac{C_F \alpha_S}{2\pi} \int\limits_{\eta}^{\tilde{x}} dx
\int\limits_{\overline{x}_{\min}(x)}^{\overline{x}^*}
d\overline{x} \, \frac{x \, \overline{x}}{2}
\mathcal{M}_{CC}\left(x,\overline{x};\, \eta\right)
\, + \,
\frac{C_F \alpha_S}{2\pi}
\int\limits_{\tilde{x}}^{x_{\max}} dx
\int\limits_{\overline{x}^{**}}^{\overline{x}^*}
d\overline{x}
\, \frac{x \, \overline{x}}{2} \mathcal{M}_{CC}\left(x,\overline{x};\,\eta\right).
\eea
The following remarks are in order:
%%%%%%%%%%%%%%%%%
\begin{enumerate}
%%%%%
\item
The second contribution to the rate
at the last member of the above equation,
involving the second root $\overline{x}^{**}$,
identically vanishes for
$\chi \ge \pi/2$.
%%%%%
\item
The first (inner) integrals, over $\overline{x}$,
in the two terms of the last member above,
are easily evaluated analytically.
The second (external) integrals, over $x$, are instead
evaluated numerically.
%%%%%
\item
As in the previous case of the quark-gluon
correlation, a cross-check of the
numerical computations
has been obtained
by fitting the EEC cumulant
and then differentiating with respect
to $\cos\chi$.
The distribution so obtained
is in complete agreement
with the direct numerical evaluation
of the differential distribution.
%%%%%%%%%%%%%%%
\end{enumerate}

%%%%%%%%%%%%%%%%%%%%%%%%%%%%%%%%%%%%%%%

\subsection{Comparison with literature}
\label{sec_Compare}

As anticipated in the Introduction, the first-order calculation of the massive EEC function was performed only numerically in the 1980s~\cite{Csikor:1983dt,Ali:1984gzn}. In this section, we present a comparison with the results obtained in those works.

In Ref.~\cite{Ali:1984gzn}, the authors evaluated the impact of heavy-quark mass effects at first order, as well as the effect of experimental resolution on the EEC function in the case of photon exchange. In Table 8 of Ref.~\cite{Ali:1984gzn}, they report the ratio between the massive and massless EEC at PETRA kinematics, $\sqrt{s}=34$ GeV. We have compared their results with the following observable:
\begin{equation}
\label{e:ratio_comp}
\mathrm{Ratio}(\chi)= \frac{1}{\sigma_0^M} \frac{d\Sigma^M}{d\cos\chi} \Bigg/ \frac{1}{\sigma_0^m} \frac{d\Sigma^m}{d\cos\chi} , ,
\end{equation}
where the superscript $M$ denotes the massive EEC, obtained by summing the contributions from Eqs.~(\ref{Qg_eval_to_num}) and~(\ref{eq_to_eval_numer}), including charm and bottom quarks with masses $m_c=1.8$ GeV and $m_b=5.0$ GeV, respectively. The superscript $m$ denotes the corresponding massless contribution at the same perturbative order.

\begin{figure}[h]
\begin{center}
\includegraphics[width=0.7\textwidth]{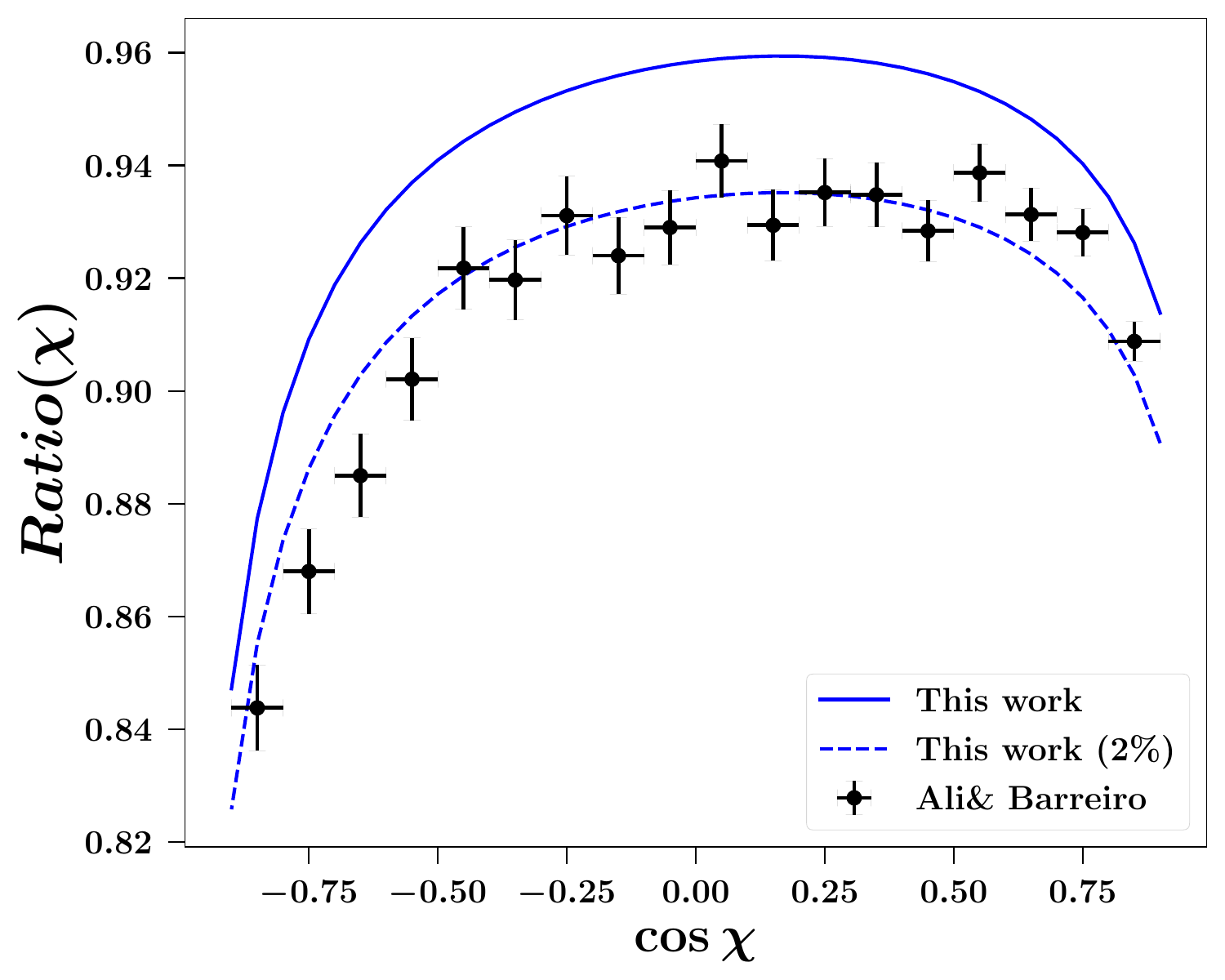}
\footnotesize
\caption{
\label{fig_comparisonAB}
\it
Comparison of the ratio between the massive and massless EEC at $\sqrt{s}=34$ GeV. The black dots correspond to the results of Ref.~\cite{Ali:1984gzn}, while the blue solid line represents our predictions. The blue dashed line shows our results shifted downward by $2\%$.
}
\end{center}
\end{figure}

In Fig.~\ref{fig_comparisonAB}, we compare the results of Ref.~\cite{Ali:1984gzn} (black dots) with our predictions (blue solid line). We observe that shifting our results downward by approximately $2\%$ leads to good agreement with the previous calculation, indicating compatibility within an estimated uncertainty of about $2\%$. A possible origin of this discrepancy may lie in the values adopted for the inclusive cross section in the normalization of the EEC in Ref.~\cite{Ali:1984gzn}, which are not fully specified.

In Ref.~\cite{Csikor:1983dt}, the author computed the massive fixed-order contribution for both photon and $Z$-boson exchange, allowing for arbitrary polarization of the $e^+e^-$ beams. In Fig.~6 of Ref.~\cite{Csikor:1983dt}, they present both the massless and massive EEC at MAC kinematics ($\sqrt{s}=30$ GeV), assuming a flavour composition proposed in Ref.~\cite{MAC}, namely events consisting of $44\%$ $b\overline{b}$, $41\%$ $c\overline{c}$, and the remaining fraction from light (effectively massless) quark pairs.

Since numerical tables are not publicly available, we digitized the curves from the original publication and constructed the corresponding ratio, following the same procedure as in the previous comparison. We therefore evaluate the observable $Ratio(\chi)$ defined in Eq.~\eqref{e:ratio_comp} for this flavour composition at $\sqrt{s}=30$ GeV, using $m_b=5$ GeV and $m_c=1.6$ GeV.

\begin{figure}[h]
\begin{center}
\includegraphics[width=0.7\textwidth]{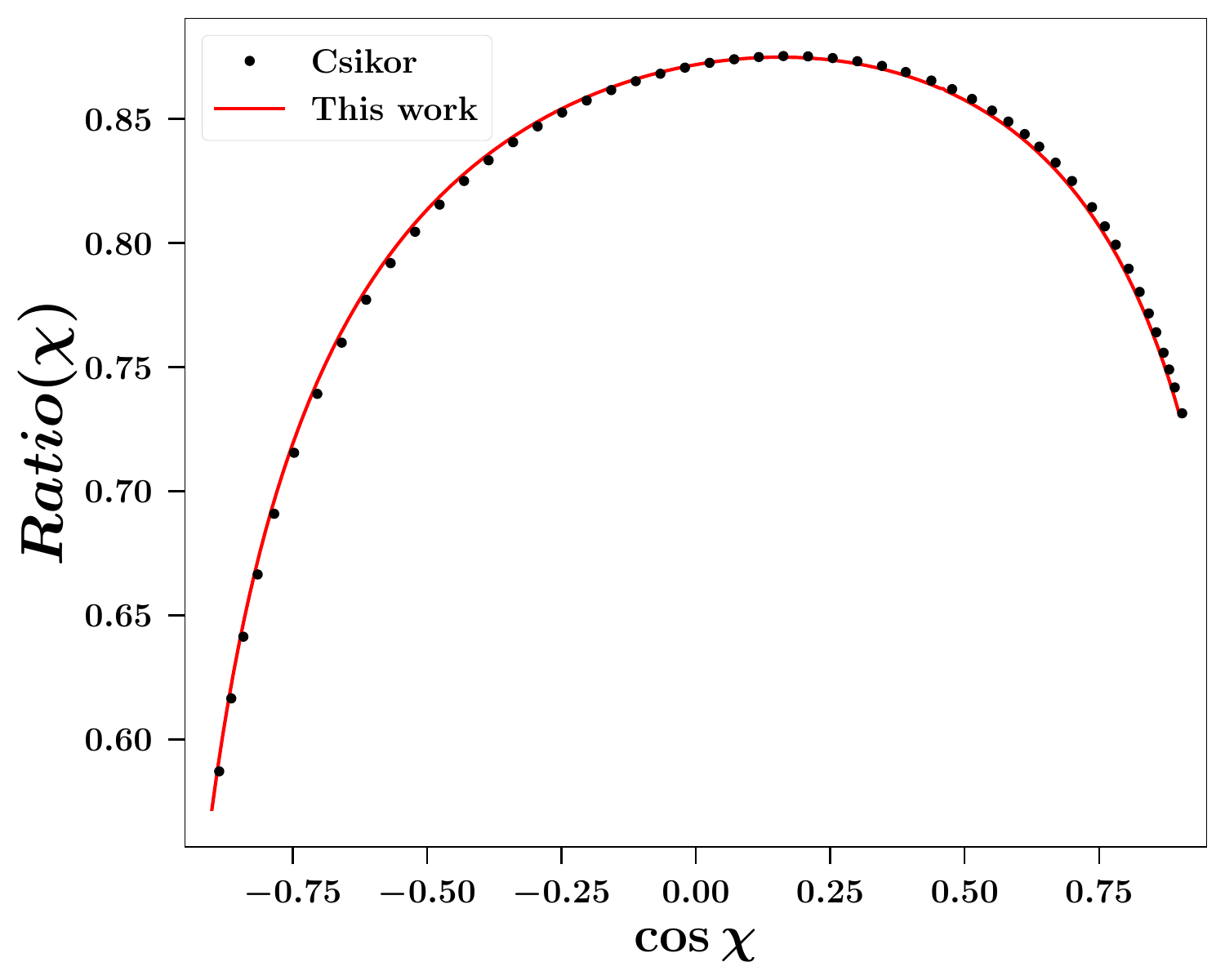}
\footnotesize
\caption{
\label{fig_comparisonCsikor}
\it
Comparison of the ratio between the massive and massless EEC at $\sqrt{s}=30$ GeV. The black dots correspond to the results of Ref.~\cite{Csikor:1983dt}, while the red solid line represents our predictions.
}
\end{center}
\end{figure}

As shown in Fig.~\ref{fig_comparisonCsikor}, we find excellent agreement with the results of Ref.~\cite{Csikor:1983dt}, both in normalization and in shape across the full angular range. A small discrepancy is observed at large values of $\cos\chi$, which may be attributed to uncertainties introduced by the digitization procedure.

Overall, from Figs.~\ref{fig_comparisonAB}--\ref{fig_comparisonCsikor}, we conclude that our results are in good agreement with the existing literature, where only numerical calculations of the massive contribution at this order were previously available.

%%%%%%%%%%%%%%%%%%%%%%%%%%%%%%%%%%%%%%%%%%%%%%%
\section{Standard massive factorization scheme}
\label{sec_stand_fact_scheme}

In this section we consider a simple
generalization  to the massive case,
of the standard factorization/resummation
scheme.
According to eq.(\ref{eq_part_ev_frac}),
the partial EEC cumulant is written
in NLL accuracy:
\bea
\label{eq_part_ev_frac_massive}
R\left( \chi; \, \chi_M ; \, \eta; \, \alpha_S\right)
&\equiv&
\frac{ \int\limits_0^{\chi}
d\Sigma/d\chi'\left( \chi'; \, \eta; \, \alpha_S \right) \, d\chi' }{ \int\limits_0^{\chi_M} d\Sigma/d\chi'\left( \chi'; \, \eta; \, \alpha_S \right) \, d\chi' } \, \simeq
\\ \nonumber
&\simeq&
\left[
1 \, + \, \frac{C_F \alpha_S}{\pi} \,
C^{(1)}(\chi_M;\eta)
\right] \,
\frac{\Phi\left( \chi; \eta; \, \alpha_S \right)}{
\Phi\left( \chi_M; \eta; \, \alpha_S \right)}
\, + \, \frac{C_F \alpha_S}{\pi} \,
\mathrm{Rem}^{(1)}\left( \chi ; \eta\right);
\eea
where $\alpha_S\equiv \alpha_S(Q)\ll 1$
and
\beq
\label{eq_CForHF_massive}
C^{(1)}(\chi_M;\eta) \, = \,
- \, \mathrm{Rem}^{(1)}\left( \chi = \chi_M ; \eta\right).
\eeq
In the massless case, the above formula
simplifies to:
\bea
\label{eq_part_ev_frac_massless}
R\left( \chi; \, \chi_M ; \, \alpha_S\right)
&\equiv&
\frac{ \int\limits_0^{\chi}
d\Sigma/d\chi'\left( \chi'; \, \alpha_S \right) \, d\chi' }{ \int\limits_0^{\chi_M} d\Sigma/d\chi'\left( \chi'; \, \alpha_S \right) \, d\chi' } \, \simeq
\\ \nonumber
&\simeq&
\left[
1 \, + \, \frac{C_F \alpha_S}{\pi} \,
C^{(1)}_0(\chi_M)
\right] \,
\frac{\Phi\left( \chi; \, \alpha_S \right)}{
\Phi\left( \chi_M; \, \alpha_S \right)}
\, + \, \frac{C_F \alpha_S}{\pi} \,
\mathrm{Rem}^{(1)}_0\left( \chi \right);
\eea
where:
\beq
C^{(1)}_0(\chi_M) \, = \,
- \, \mathrm{Rem}^{(1)}_0\left(\chi = \chi_M\right),
\eeq
with:
\beq
\mathrm{Rem}^{(1)}_0\left(\chi\right)
\, \equiv \,
\mathrm{Rem}^{(1)}\left(\chi; \, \eta = 0 \right).
\eeq
The main task is then to explicitly evaluate
the first-order massive Remainder function.

%%%%%%%%%%%%%%%%%%%%%%%%%%%%%%%%%%%%%%%%

\subsection{Massive remainder functions}
\label{sec_mass_rem_fun}

In this section we evaluate numerically
the leading-order remainder functions in the massive
case, for the factorized EEC differential
distribution.
We work at first order in perturbation theory,
i.e. at $\mathcal{O}(\alpha_S)$.
According to eq.(\ref{eq_EEC_diff_fact}),
one has to subtract, from the massive EEC spectrum
(computed in the previous sections),
the expansion to first order of the massive Sudakov form factor:
\beq
\frac{C_F \alpha_S}{2\pi} \,
\mathrm{rem}^{(1)}(\chi;\eta)
\, \equiv \,
\frac{1}{\sigma}
\, \frac{d\Sigma}{d\chi}(\chi;\eta;\alpha_S)
\, - \,
\left[
\varphi(\chi; \eta; \alpha_S)\right]_{O(\alpha_S)}
\qquad
(0 < \chi < \pi).
\eeq

%%%%%%%%%%%%%%%%%%%%%%%%%%%%%%%%%%%%%%%%%%%

\subsubsection{Massive Sudakov form factor}

By expanding up to first order the resummed massive Sudakov form factor derived by two of us in
\cite{Aglietti:2024zhg},
one obtains in the bulk
($0<\chi<\pi$):
\bea
\label{eq_main_esp_disc}
\varphi(\chi; \eta; \alpha_S)
&=& \frac{C_F \, \alpha_S}{2\pi}
(-)\cot\left(\frac{\chi}{2}\right)
\Bigg\{
\left[
2 \log \left( \sin \left( \frac{\chi}{2} \right) \right)
\, + \, \frac{3}{2}
\right]
\theta\left[ \chi \, - \, 2 \arcsin\left(\frac{\eta}{2}\right) \right] + \,\,\,\,\quad
\\ \nonumber
&& \qquad\qquad\qquad\qquad\, + \, \left[
2 \log\left(\frac{\eta}{2}\right) \, + \, 1
\right]
\theta\left[ 2 \arcsin\left(\frac{\eta}{2}\right)
\, - \, \chi \right]
\Bigg\} \, + \, \mathcal{O}(\alpha_S^2).
\eea
Eq.(\ref{eq_main_esp_disc}) is obtained by
considering the integrand of the radiator (i.e. the exponent) of the massive Sudakov in $b$-space,
removing the factor $\left[J_0(b k_\perp)-1\right]$,
and making inside it the replacement
\beq
\label{eq_final_repl}
k_\perp \, \mapsto \, Q \, \sin\left(\frac{\chi}{2}\right).
\eeq
The above rule comes from the fact that
the (direct) Fourier-Bessel transform involves
the conjugate variables
\beq
k_\perp \, \mapsto \, b,
\eeq
while the inverse transform involves
\beq
b \, \mapsto \, Q \, \sin\left(\frac{\chi}{2}\right).
\eeq
By composing the direct with the inverse transform, one obtains the
identity transformation with
the change of variable (\ref{eq_final_repl}).

The following remarks are in order:
%%%%%%%%%%%%%%%%%
\begin{enumerate}
%%%%%
\item
Since the form factor only
describes soft and/or collinear radiation,
it is a good approximation of the complete
distribution only for
\beq
0 \, < \, \chi \, \ll \, 1.
\eeq
We have checked that our numerical distributions
behave like the form factor in the above
region.
%%%%%
\item
The $\theta$-functions above
basically define an {\it effective massless
region}:
\beq
\chi \, > \, 2 \arcsin\left(\frac{\eta}{2}\right)
\, \simeq \, \eta
\qquad (\eta \, \ll \, 1),
\eeq
and a {\it substantially massive one}:
\beq
\chi \, < \, 2 \arcsin\left(\frac{\eta}{2}\right).
\eeq
%%%%%
\item
Eq.(\ref{eq_main_esp_disc}) has a finite discontinuity
related to the subleading, single-logarithmic terms, which can be
regulated by replacing the step function
for example with a sigmoid:
\beq
\theta(x)
\quad
\mapsto
\quad
\theta_\sigma(x) \, \equiv \,
\frac{1}{\sqrt{2\pi\sigma^2}}
\int\limits_{-\infty}^x e^{-(y-x)^2/(2\sigma^2)} dy
\qquad (\sigma > 0).
\eeq
In a weak sense indeed:
\beq
\lim_{\sigma\to 0^+} \theta_\sigma(x)
\, = \, \theta(x).
\eeq
A good regularization is then obtained
by taking $0 <\sigma \ll 1 $
(in the resummed form factor, a regularization
is {\it de facto} provided by the Fourier-Bessel
transform ($k_\perp\mapsto b$) of the one-gluon distribution
in $k_\perp$-space).
%%%%%%%%%%%%%%%
\end{enumerate}

%%%%%%%%%%%%%%%%%%%%%%%%%%%%%%%%%%%%%%%%%%%%%%%%%

\subsubsection{Improved massive Sudakov form factor}

In addition to the discontinuity cited
above, the form factor $\varphi(\chi,\eta,\alpha_S)$ incorporates
mass effects in the ultra-relativistic
approximation $0 < m/Q \ll 1$ only.
One can construct an improved form
factor which does not possess the
undesirable properties above,
i.e. a form factor which is smooth in
the variable $\chi$
and incorporates mass effects
for any value of the ratio $m/Q<1/2$
\cite{noilavorofuturo}.
Such improved massive Sudakov form factor
$\varphi_{\mathrm{imp}}$ has the following first-order expansion:
\bea
\label{main_eq_EEC_soft}
\varphi_{\mathrm{imp}}\left(\chi;\eta;\alpha_S\right)
&=& \frac{C_F \alpha_S}{2\pi}
\, \frac{1}{2}
\cot\left(\frac{\chi}{2}\right)
\Bigg\{
\frac{1 \, + \, u^2}{u}
\log\left[ \frac{1 \, + \, u}{ 1 \, - \, u \, + \,
u\left(1-\cos\chi\right)/z_{\max}^2 } \right] +
\nonumber\\
&& \quad\qquad\qquad\qquad\,\,\,
- \, \frac{1 \, - \, u^2}{u}
\left[
\frac{1}{ 1 \, - \, u \, + \, u
\left( 1 \, - \, \cos\chi \right)/z_{\max}^2 }
\, - \, \frac{1}{1 \, + \, u}
\right]
\Bigg\} +
\nonumber\\
&+& \frac{C_F \, \alpha_S}{2\pi}
\left( - \, \frac{3}{2} \right)
\frac{\sin\left( \chi/2 \right) \,
\cos\left(\chi/2\right)}{
\sin^2\left(\chi/2\right)
\, + \, \eta^2/4}
\,\, + \,\, \mathcal{O}\left(\alpha_S^2\right);
\eea
where:
%%%%%%%%%%%%%%%%%
\begin{enumerate}
%%%%%
\item
$u\equiv ds/dt$ is the kinematic heavy quark 3-velocity in the soft limit
in the C.O.M. frame,
\beq
u \, = \, \sqrt{1 \, - \, \eta^2}.
\eeq
\item
$z_{\max}$ is the maximal effective
soft-gluon energy, below which
the angular integration is unconstrained
($0 \le \theta_{Qg} \le \pi$),
\beq
z_{\max} \, = \, \frac{1 \, - \, \eta}{1 \, - \, \eta/2}.
\eeq
\end{enumerate}
The following remarks are in order.
%%%%%%%%%%%%%%%%%
\begin{enumerate}
%%%%%
\item
The form factor on the r.h.s. of eq.(\ref{main_eq_EEC_soft})
has the correct massless limit:
\beq
\lim_{\eta\to 0^+} \varphi_{\mathrm{imp}}
\, = \,
\frac{C_F \, \alpha_S}{2\pi}
(-)\cot\left(\frac{\chi}{2}\right)
\left\{
2 \log \left[ \sin \left( \frac{\chi}{2} \right) \right]
\, + \, \frac{3}{2}
\right\}.
\eeq
The expression above coincides indeed with the formula on the first line on the r.h.s. of eq.(\ref{eq_main_esp_disc}).
Note that the condition
\beq
\sin\left(\frac{\chi}{2}\right)
\, > \, \frac{\eta}{2}
\eeq
is equivalent to the condition
\beq
\chi \, > \, 2
\arcsin\left(\frac{\eta}{2}\right).
\eeq
%%%%%
\item
The form factor on the r.h.s. of eq.(\ref{main_eq_EEC_soft})
has the correct Ultra-Relativistic
($\mathrm{UR}$) limit.
By assuming indeed:
\beq
0 \, < \, \chi \, \ll \, \eta \, \ll \, 1,
\eeq
one obtains at leading order in $\eta$:
\beq
\left[
\varphi_{\mathrm{imp}}(\chi,\eta,\alpha_S)\right]_{\mathcal{O}(\alpha_S),\mathrm{UR}}
\, = \,
\frac{C_F \, \alpha_S}{2\pi} \,
(-)\cot\left(\frac{\chi}{2}\right)
\left[
2 \log\left(\frac{\eta}{2}\right) \, + \, 1
\right].
\eeq
The above expression is in complete
agreement with the formula
on the second line on the r.h.s. of
eq.(\ref{eq_main_esp_disc}).
%%%%%%%%%%%%%%%
\end{enumerate}
The remainder function in the new
scheme, containing the improved
$(\mathrm{imp})$ form factor 
$\varphi_{\mathrm{imp}}$,
is then given by:
\beq
\frac{C_F \alpha_S}{2\pi} \,
\mathrm{rem}^{(1)}_{\mathrm{imp}}(\chi,\eta,\alpha_S)
\, \equiv \,
\frac{1}{\sigma}
\, \frac{d\Sigma}{d\chi}(\chi,\eta,\alpha_S)
\, - \,
\left[\varphi_{\mathrm{imp}}(\chi,\eta,\alpha_S) \right]_{\mathcal{O}(\alpha_S)}.
\eeq
It is trivial to calculate the
remainder function in one scheme,
once it is known in the other one:
\beq
\frac{C_F \, \alpha_S}{2\pi} \,
\mathrm{rem}^{(1)}_{\mathrm{imp}}(\chi,\eta,\alpha_S)
\, = \,
\frac{C_F \, \alpha_S}{2\pi} \,
\mathrm{rem}^{(1)}(\chi,\eta,\alpha_S)
\, + \,
\left[\varphi(\chi,\eta,\alpha_S) \right]_{\mathcal{O}(\alpha_S)}
\, - \,
\left[ \varphi_{\mathrm{imp}}(\chi,\alpha_S) \right]_{\mathcal{O} (\alpha_S)},
\eeq
and vice versa.
It is also trivial to re-obtain the complete
distribution from anyone of the remainder
functions:
\bea
\frac{1}{\sigma}
\, \frac{d\Sigma}{d\chi}(\chi,\eta,\alpha_S)
&=&
\frac{C_F \, \alpha_S}{2\pi} \,
\mathrm{rem}^{(1)}(\chi,\eta,\alpha_S)
\, + \,
\left[ \varphi(\chi,\eta,\alpha_S) \right]_{\mathcal{O}(\alpha_S)} \, =
\nonumber\\
&=&
\frac{C_F \, \alpha_S}{2\pi} \,
\mathrm{rem}^{(1)}_{\mathrm{imp}}(\chi,\eta,\alpha_S)
\, + \,
\left[ \varphi_{\mathrm{imp}}(\chi,\eta,\alpha_S) \right]_{\mathcal{O}(\alpha_S)}.
\eea
From now on, we will consider
remainder functions in the improved
scheme only (Sudakov = $\varphi_{\mathrm{imp}}$) and,
to simplify notation, the subscript
``$\mathrm{imp}$'' will be omitted.

%%%%%%%%%%%%%%%%%%%%%%%%%%%%%%%%%%%%%%%

\subsection{Comparing different masses}

\begin{figure}[ht]
\begin{center}
\includegraphics[width=0.9\textwidth]{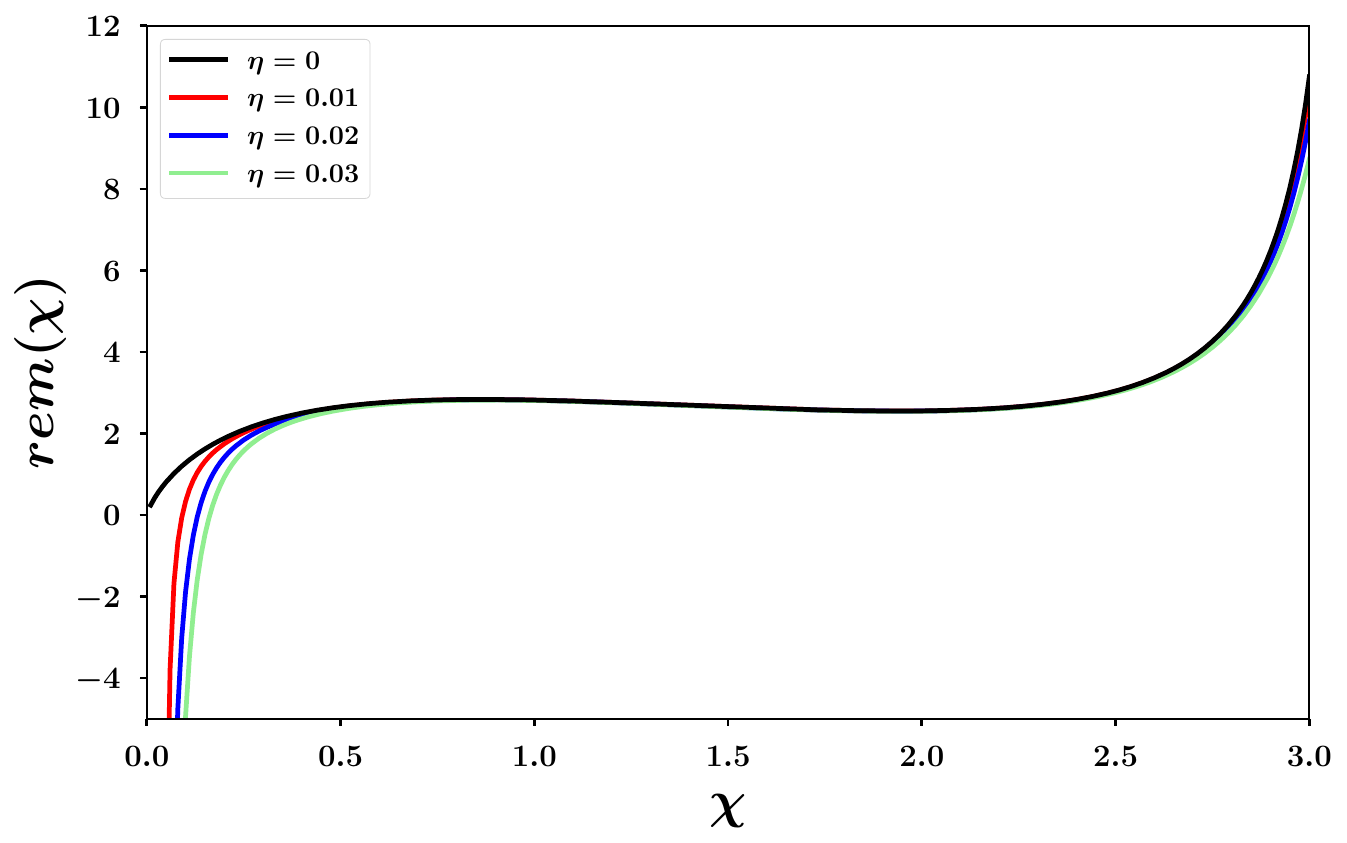}
\footnotesize
\caption{
\label{fig_plotremcomp}
\it
First-order remainder functions for the
EEC spectrum in the massless case $\eta=0$
(black line) and in almost massless cases
(red line: $\eta=0.01$;
blue line: $\eta=0.02$;
green line: $\eta=0.03$).
}
\end{center}
\end{figure}

In this section, we consider remainder functions,
evaluated as described in the previous section,
for different values of the heavy-quark mass
parameter $\eta$.
Let us begin comparing the remainder functions of the EEC spectrum in the
{\it massless} case $\eta=0$ and
in the {\it almost-massless} cases
$\eta=0.01, \, 0.02$ and $\eta=0.03$%
%%%%%%%%%%
\footnote{
The smallest value of the mass parameter we have chosen, $\eta=0.01$, corresponds, for example, to a quark
with a mass $m \simeq 0.5\,\mathrm{GeV}$
--- let us say a strange quark with a constituent
mass --- at the $Z^0$ peak ($Q=m_Z=91.2\,\mathrm{GeV}$).
The value $\eta=0.03$ correspond instead
to a charm quark ($m_c \simeq 1.5\,\mathrm{GeV}$),
still at the $Z^0$ peak.
}
%%%%%%%%%%
(see fig.$\,$\ref{fig_plotremcomp}).
We observe that the curves are barely distinguishable from each other
for $\chi \gsim 0.1\div 0.15$; below that value,
the massless curve approaches zero%
%%%%%%%%%%
\footnote{
The remainder function of the
distribution in $\cos\chi$,
i.e. $d\Sigma/d\cos\chi$,
diverges instead logarithmically
for $\chi\to 0^+$,
as is usually the case.
},
%%%%%%%%
while the massive ones go down
quite fast to $-\infty$,
i.e. they have a downwards
vertical asymptote at $\chi=0^+$;
these singularities however turn out
to be integrable (as it should), as far
as an extrapolated numerical computation can tell.
We have evaluated numerically the EEC
remainder functions down to
$\chi = \chi_{\min}=10^{-3}$,
corresponding to
$y \equiv (1-\cos\chi)/2 \simeq 2.5 \times 10^{-7}$.

%
%%%%%%%%%%%%%%%%%%%
%
%
%%%%%%%%%%%
%

%%%%%%%%%%%%%%%%%%%%%%%%%%%%%%%%%%%%%%%%%%%%%%%

\subsection{Partially-integrated distributions}

Let us now consider the Remainder
functions of the partially-integrated
distributions,
which are trivially obtained
by integrating the
remainder functions of the EEC spectrum.

The partially-integrated Remainder function
can be evaluated in the massless case,
unlike the massive one,
in closed exact analytic form.
By removing the usual overall factor
$C_F \alpha_S/(2\pi)$, it reads:
\bea
\label{eq_Rem0_analit}
\mathrm{Rem}^{(1)}_0(\chi) &=&
\, + \, \frac{72 \log \left[\sin\left(\chi/2\right)\right]}{\left[ \cos(\chi) \, + \, 1 \right]^4}
\, - \, \frac{\cos (2 \chi) + 320 \log\left[
\sin\left(\chi/2 \right)\right] - 73}{
4 \left[ \cos (\chi) \, + \, 1 \right]^3} +
\nonumber\\
&& + \, 3 \, \frac{8 \log \left[\sin \left(\chi/2\right)\right]-11}{
2\left[\cos (\chi) \, + \, 1\right]^2}
\, - \, \frac{4 \log \left[\sin\left(\chi/2\right)\right]}{\cos(\chi) \, + \, 1} \, +
\\ \nonumber
&&
+ \, \frac{1}{24} \left\{ - \, 24 \, \text{Li}_2\left[\cos^2\left(\frac{\chi}{2}\right)\right]
\, + \, 108 \log\left[\sin \left(\frac{\chi}{2}\right)\right]
\right. +
\nonumber\\
&& \qquad\qquad\qquad\qquad
 - \, \left. 36 \log
\left[\cos\left(\frac{\chi}{2}\right)\right]
+ 4 \pi^2 + 45 \right\}
\qquad\qquad
(0<\chi<\pi).
\nonumber
\eea
It is immediate to check that
it vanishes in the two-jet limit,
as it should (cf. eq.(\ref{eq_imp_prop_Rem_fun})):
\beq
\lim_{\chi\to 0^+} \mathrm{Rem}^{(1)}_0(\chi) \, = \, 0.
\eeq
It is a strictly-monotonically-increasing
function of $\chi$ for
$\chi \in (0,\pi)$
and it diverges as $1/(\pi-\chi)$
for $\chi \to \pi^-$
because of hard collinear effects
occurring in the forward region.
Let us give some typical values
of the massless Remainder function:
\beq
\mathrm{Rem}^{(1)}_0\left(\chi=\frac{\pi}{2}\right) \, = \,
\frac{1}{24} \left[ 93+2 \pi ^2+12 (\log 2-3)
\log 2 \right]
\, = \, 3.89797\cdots;
\eeq
while:
\beq
\mathrm{Rem}^{(1)}_0\left(\chi=2.5\right)
\, = \,
6.35350\cdots.
\eeq
A good fit of the massless remainder function
in the region
\beq
0 \, < \, \chi \, \lsim \, 3
\eeq
is provided by the following fifth-order polynomial
in $\chi$:
\beq
\mathrm{Rem}^{(1)}_0(\chi) \, \simeq \,
- \, 0.0171482 \, \chi^5
\, + \, 0.390459 \, \chi^4
\, - \, 1.8177 \, \chi^3
\, + \, 3.12535 \, \chi^2
\, + \, 0.660832 \, \chi.
\eeq
The property discussed in the previous
section implies that the quasi-massless
$(\eta \le 0.03)$
Remainder functions are simply shifted downwards
with respect to the massless one
for $\chi \gsim 0.1\div 0.15$.
A detailed analysis shows that
the small-mass Remainder
functions can be written, to a very good approximation
($\mathcal{O}(1\%)$, let's say),
in a wide $\chi$-interval,
in the following way:
\beq
\label{eq_Rem_new}
\mathrm{Rem}^{(1)}\left(\chi; \, \eta \right)
\, \simeq \,
\mathrm{Rem}^{(1)}_0(\chi)
\, - \, a_\eta \left( 1 \, - \, e^{ - \, b_\eta \, \chi} \right),
\qquad 0 < \chi \lsim 3,
\eeq
where $\mathrm{Rem}^{(1)}_0(\chi)$ is the {\it massless} Remainder function
discussed above,
\beq
\mathrm{Rem}^{(1)}_0\left(\chi\right) \, \equiv \,
\mathrm{Rem}^{(1)}\left(\chi; \, \eta = 0 \right).
\eeq
The quantities
$a_\eta>0$ and $b_\eta>0$ are two $\eta$-dependent
numerical coefficients.
Actually, we have found that
eq.(\ref{eq_Rem_new}) gives a good
fit of the massive Remainder functions
in both the vector and the axial cases,
up to surprisingly large values of the
mass parameter, namely up to $\eta \simeq 0.5$.
However, by increasing $\eta$,
the good-fit region $[0,\eta_{\max}]$ shrinks a bit
and one has to take
lower values of $\eta_{\max}$.

\begin{table}[htbp]
\begin{center}
\begin{tabular}{|l|l|l|}
\hline
$\eta$ & $a_{\eta}$ & $c_{\eta}$
\\
\hline
0.01 & 6.57328 & 1.28693
\\
\hline
0.02 & 6.48409 & 1.32675
\\
\hline
0.03 & 6.42179 & 1.35580
\\
\hline
0.06 & 6.24206 & 1.41287
\\
\hline
0.10 & 6.08681 & 1.47943
\\
\hline
0.20 & 5.72950 & 1.59345
\\
\hline
0.30 & 5.34508 & 1.64955
\\
\hline
0.40 & 4.97800 & 1.68793
\\
\hline
0.50 & 4.57078 & 1.70782
\\
\hline
\end{tabular}
\end{center}
\caption{Numerical values of the coefficients
$a_\eta$ and $c_\eta$
in the case of the {\it vector} current.}
\label{table1}
\end{table}

\begin{table}[htbp]
\begin{center}
\begin{tabular}{|l|l|l|}
\hline
$\eta$ & $a_{\eta}$ & $c_{\eta}$
\\
\hline
0.01 & 6.57260 & 1.28727
\\
0.02 & 6.48183 & 1.32781
\\
0.03 & 6.41735 & 1.35782
\\
0.06 & 6.26916 & 1.42813
\\
0.10 & 6.09009 & 1.50045
\\
0.20 & 5.66966 & 1.63089
\\
0.30 & 5.21221 & 1.69785
\\
0.40 & 4.79701 & 1.75289
\\
0.50 & 4.37046 & 1.80415
\\
\hline
\end{tabular}
\end{center}
\caption{Numerical values of the coefficients
$a_\eta$ and $c_\eta$
in the case of the {\it axial} current.}
\label{table2}
\end{table}

\begin{figure}[ht]
\begin{center}
\includegraphics[width=0.9\textwidth]{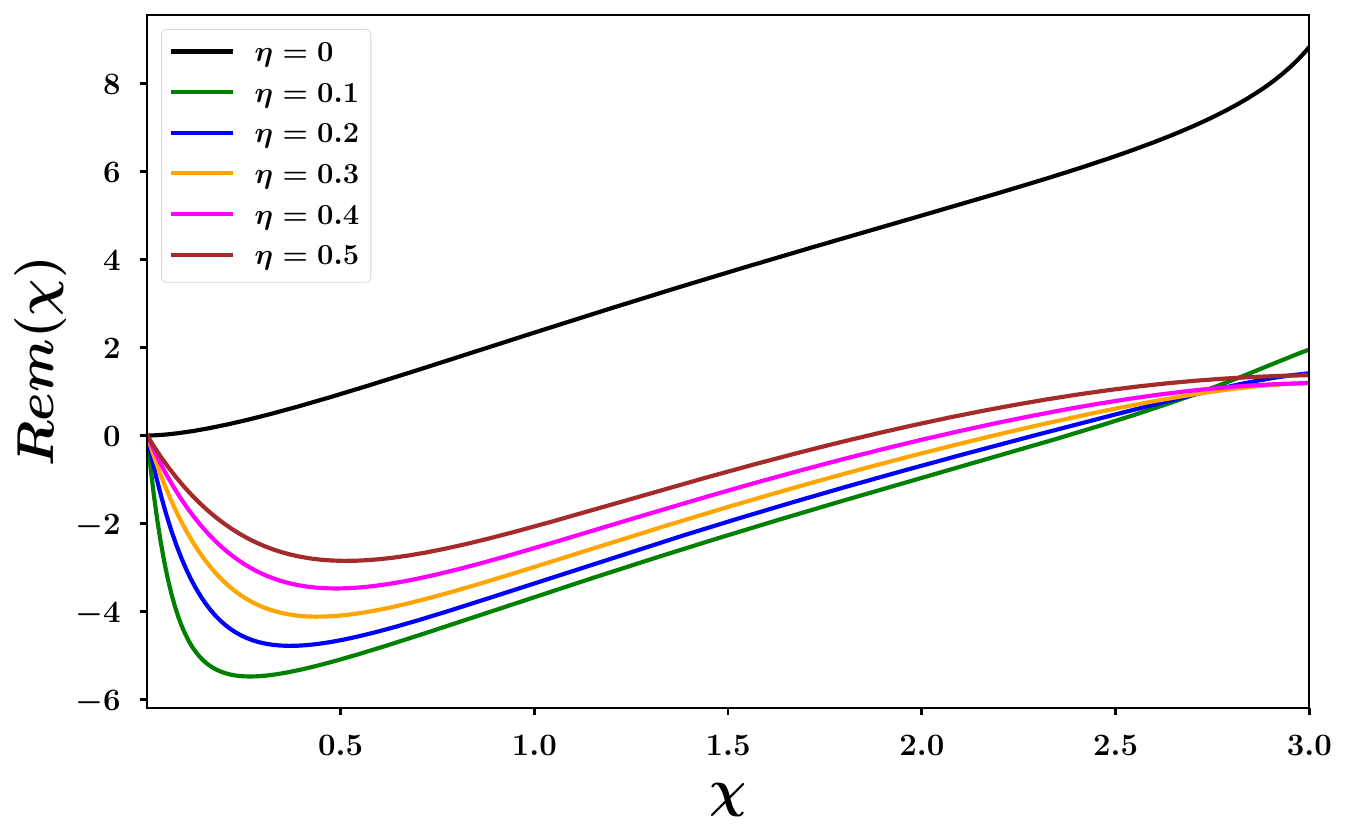}
\footnotesize
\caption{
\label{fig_plotRemfunc}
\it
First-order Remainder functions for the
partially-integrated EEC distribution
in the vector case (the axial case is similar).
Black line: $\eta=0$ (massless case);
green: $\eta=0.1$ (beauty at LEP1);
blue: $\eta=0.2$;
orange: $\eta=0.3$;
magenta: $\eta=0.4$;
brown: $\eta=0.5$.
}
\end{center}
\end{figure}

Let us comment on these coefficients:
%%%%%%%%%%%%%%%%%
\begin{enumerate}
%%%%%
\item
The coefficient $a_\eta \sim 4.5\div 6.5$ has a mild dependence on $\eta$; its values for our
choices of $\eta$ are collected in tables \ref{table1} and \ref{table2}
for both the vector and axial cases
respectively;
%%%%%
\item
The coefficient $b_\eta$ is instead roughly
inversely proportional to $\eta$, so it is
natural to write it as:
\beq
b_\eta \, = \, \frac{c_\eta}{\eta}.
\eeq
The coefficient $c_\eta$, expected to be of order one, has a mild dependence
on $\eta$; its values are also reported in
tables \ref{table1} and \ref{table2}.
As far as a numerical calculation can tell,
we find that:
\beq
\lim_{\eta \to 0^+} b_\eta \, = \, + \, \infty.
\eeq
%%%%%
\item
In order to accurately extract
$a_\eta$ and $b_\eta$, in the small
mass cases, namely
for $\eta < 0.1$,
we have extrapolated the remainder
functions up to $\chi=0^+$.
%%%%%%%%%%%%%%%
\end{enumerate}
The massive (and massless) Remainder functions for the partially-integrated distribution are shown in fig.~\ref{fig_plotRemfunc}, for several values of the mass parameter $\eta$.

Eq.(\ref{eq_Rem_new}) is one of the main
results of our paper.
It says that the dependence on the heavy-quark mass
parameter $\eta$
is totally relegated in the second term of
its r.h.s..
Furthermore, the massive correction to
the massless remainder correction is negative.

According to eq.(\ref{eq_CForHF_massive}), the first-order massive coefficient function
is given by:
\bea
C^{(1)}\left(\chi_M; \, \eta\right) &=&
- \, \mathrm{Rem}^{(1)} \left(\chi_M; \, \eta\right) \, =
\nonumber\\
&=& - \, \mathrm{Rem}^{(1)}_0 \left(\chi_M\right)
\, + \, a_\eta \left(
1 \, - \, e^{- \, b_\eta \, \chi_M} \right) \, =
\nonumber\\
&=& \,\,\,\,\, C^{(1)}_0\left(\chi_M\right)
\, + \, a_\eta \left(
1 \, - \, e^{- \, b_\eta \, \chi_M} \right).
\eea
The last equality above says that the massive coefficient function has a positive correction with respect
to the massless one
$C^{(1)}_0\left(\chi_M\right)$,
which is of order $a_\eta$, as
$b_\eta \, \chi_M \gsim 1$.

%
%%%%%%%%%%%%%%%%%%%
%
%
%%%%%%%%%%%
%

%%%%%%%%%%%%%%%%%%%%%%%%%%%%%%%%%%%%%%%%%%%%%%%%%%%%%%%%%%%%%%

\section{New massive factorization scheme with a smooth massless limit}
\label{sec_new_massive_factor}

The massless limit of the
mass-dependent term on the r.h.s. of
eq.(\ref{eq_Rem_new}) is not vanishing:
\beq
\lim_{\eta\to 0^+} - \, a_\eta
\left( 1 \, - \, e^{ - \, b_\eta \, \chi} \right)
\, = \,  - \, a_\eta \, \ne \, 0
\qquad (\chi>0).
\eeq
That implies that the massless limit
of the massive Remainder function given in
eq.(\ref{eq_Rem_new}) is not equal
to the {\it ab initio} massless Remainder
function (obtained from the factorization
of the fixed-order massless distribution):
\beq
\lim_{\eta\to 0^+}
\mathrm{Rem}^{(1)}\left(\chi; \eta \right)
\, = \,
\mathrm{Rem}^{(1)}_0(\chi)
\, - \, a_\eta
\, \ne \,
\mathrm{Rem}^{(1)}_0(\chi).
\eeq
The limit Remainder function
is shifted downwards by
the quite large amount $a_\eta \sim 6.5$
with respect to the {\it ab initio} massless one.
Such discontinuity is not reasonable on
physical ground, because one expects a massive
Remainder function to be close to
the massless one for very small masses.
Furthermore, the massless limit
of the massive Remainder does not
vanish in the two-jet limit $\chi\to 0^+$,
as one would expect.
Moreover, perturbative fixed-order calculations
of the EEC distribution
do not exhibit any discontinuity in the massless limit.
Therefore, the discontinuity phenomenon we have found above,
is to be considered
as an artifact of the standard factorization scheme,
once it is generalized in a straightforward way to the massive case.
In the latter case, we are dealing indeed with
a truly multi-scale process.

The basic point is that, in the mass-dependent term, the massless limit $\eta\to 0^+$
and the two-jet limit $\chi \to 0^+$
do not commute with each other, as:
\bea
\lim_{\eta \to 0^+} \left[ \lim_{\chi\to 0^+}
- \, a_\eta \left( 1 \, - \, e^{-b_\eta \chi} \right)
\right]
&=& 0;
\nonumber\\
\lim_{\chi\to 0^+} \left[ \lim_{\eta\to 0^+}
- \, a_\eta \left( 1 \, - \, e^{-b_\eta \chi} \right)
\right]
&=& - \, a_\eta.
\eea
However, an improved massive factorization scheme,
involving a Remainder function with a
smooth massless limit,
can be constructed, by noticing that
both the massless Remainder function and
the mass-dependent term in eq.(\ref{eq_Rem_new}) do vanish in the two-jet limit $\chi \to 0^+$.
Therefore one can ``move'' the mass-dependent term
from the
massive Remainder function
$\mathrm{Rem}^{(1)}(\chi; \, \eta)$
to the massive first-order coefficient
function $C^{(1)}=C^{(1)}(\chi_M; \, \eta)$.
Let us remark that,
since the coefficient function is multiplied
by the Sudakov ratio $\Phi/\Phi_M=1+\mathcal{O}(\alpha_S)$,
such transformation does not modify
the expansion up to first order
of the resummed distribution.

One then defines a new,
$\chi$-dependent first-order
coefficient function
$\tilde{C}^{(1)} = \tilde{C}^{(1)}(\chi_M; \eta; \chi)$ as follows:
\bea
\label{eq_def_new_CF}
\tilde{C}^{(1)}\left(\chi_M; \eta; \, \chi \right)
&\equiv&
C^{(1)}\left( \chi_M; \, \eta \right)
\, - \, a_\eta \left( 1 \, - \,
e^{- \, b_\eta \, \chi}
\right) \, =
\nonumber\\
&=& - \,
\mathrm{Rem}^{(1)}\left(\chi_M; \, \eta \right)
\, - \, a_\eta
\left( 1 \, - \, e^{- \,b_\eta \, \chi} \right) \, =
\nonumber\\
&=& - \, \mathrm{Rem}^{(1)}_0\left(\chi_M\right)
\, + \, a_\eta \left(
e^{ - \, b_\eta \, \chi}
\, - \, e^{ - \, b_\eta \, \chi_M}
\right).
\eea
In the second equality above
(the first equation is a definition), we have used
eq.(\ref{eq_unitary_C_Rem}),
while in the third one
we have used eq.(\ref{eq_Rem_new}).
Since the {\it ab initio} massless coefficient function is given by:
\beq
C^{(1)}_0\left( \chi_M \right)
\, = \, - \,
\mathrm{Rem}^{(1)}_0\left(\chi_M\right),
\eeq
the new coefficient function can be written
in terms of the massless one
in the ``suggestive'' form:
\beq
\label{eq_suggestive}
\tilde{C}^{(1)}\left(\eta,\chi_M; \, \chi \right)
\, = \, C^{(1)}_0\left( \chi_M \right)
\, + \, a_\eta \left(
e^{ - \, b_\eta \, \chi}
\, - \, e^{ - \, b_\eta \, \chi_M}
\right).
\eeq
Having removed the mass-dependent term
from the massive Remainder function,
the latter,
in the improved tilde ($\tilde{\,}$) scheme, is simply equal
to the massless one:
\beq
\tilde{\mathrm{Rem}}^{(1)}\left(\chi; \eta \right)
\, = \,
\mathrm{Rem}^{(1)}_0(\chi).
\eeq

The following remarks are in order.
%%%%%%%%%%%%%%%%%
\begin{enumerate}
%%%%%
\item
Within the approximations we are working
in, the new factorization scheme,
non-logarithmic mass effects are completely relegated
in the coefficient function
(as well as logarithmic mass effects
were relegated in the Sudakov form factor only \cite{Aglietti:2024zhg});
they completely disappeared
from the Remainder function.
%%%%%
\item
According to eq.(\ref{eq_suggestive}),
the improved (massive) coefficient function
converges to the massless one
in the massless limit:
\beq
\lim_{\eta \to 0^+}
\tilde{C}^{(1)}\left(\eta,\chi_M; \, \chi \right)
\, = \, C^{(1)}_0\left( \chi_M \right)
\qquad (\chi>0).
\eeq
The improved scheme then has a smooth massless
limit, as requested.
%%%%%
\item
Since $0\le\chi \le \chi_M$,
according to eq.(\ref{eq_suggestive}),
the new massive
coefficient function $\tilde{C}^{(1)}$
has, in general, a positive correction
with respect to the massless one
$C^{(1)}_0$, which is given by
\beq
a_\eta \left(
e^{ - \, b_\eta \, \chi}
\, - \, e^{ - \, b_\eta \, \chi_M}
\right) \, \ge \, 0
\qquad
(0 \, \le \, \chi \, \le \, \chi_M).
\eeq
The above correction becomes large --- of order $a_\eta$ ---
in the two-jet region
$b_\eta \, \chi \lsim 1$.
It follows that, in general, the new coefficient function,
\beq
\tilde{C}(\chi,\alpha_S) \, \simeq \,
1 \, + \, \frac{C_F \alpha_S}{\pi} \, \tilde{C}^{(1)}(\chi),
\eeq
is larger
in the two-jet region than the massless one,
\beq
C(\alpha_S)\, \simeq \,
1 \, + \, \frac{C_F \alpha_S}{\pi} \, C^{(1)}.
\eeq
Since, in the partial EEC cumulant, the coefficient function
multiplies the Sudakov form factor ratio
$\Phi/\Phi_M$,
we expect the mass effects
under consideration
to produce an {\it enhancement}
of the predicted rate
in the two-jet region.
%%%%%
\item
According to the first equality in eq.(\ref{eq_def_new_CF}),
the improved massive coefficient function
has a negative correction,
vanishing at $\chi=0$, with respect
to the standard massive one.
%%%%%
\item
A similar non-commutativity problem
to the present one has also been found in threshold
resummation in the massive case
\cite{Aglietti:2022rcm,Gaggero:2022hmv,Ghira:2023bxr,Ghira:2024nkk}.
A similar solution, involving a new
factorization scheme, has also been found
in that case.
%%%%%
\item
By differentiating with respect to
$\chi$ the partial EEC cumulant,
one obtains for the differential
distribution:
\beq
\frac{1}{Z} \, \frac{1}{\sigma}
\frac{d\Sigma}{d\chi}(\chi)
\, = \, \tilde{C}(\chi) \,
\frac{\varphi(\chi)}{\Phi_M} \, + \,
\frac{d\tilde{C}}{d\chi}(\chi)
\frac{\Phi(\chi)}{\Phi_M}
\, + \, \frac{C_F \alpha_S}{\pi}
\, \mathrm{rem}^{(1)}(\chi);
\eeq
where:
\beq
Z \, \equiv \, \frac{1}{\sigma} \, \int\limits_0^{\chi_M}
\frac{d\Sigma}{d\chi}\, d \chi
\, = \, \frac{1}{2} \, + \,
\mathcal{O}\left(\alpha_S\right)
\qquad \left(\chi_M < \pi\right).
\eeq
The derivative of the improved
coefficient function with respect
to the correlation angle $\chi$
explicitly reads:
\beq
\frac{d\tilde{C}}{d\chi}(\chi)
\, = \, \frac{C_F \alpha_S}{\pi}
\, (-) \, a_\eta \, b_\eta \, e^{\, - \, b_\eta \chi}
\, + \, \mathcal{O}\left(\alpha_S^2\right).
\eeq
Note that it is negative.
%%%%%%%%%%%%%%%
\end{enumerate}

%%%%%%%%%%%%%%%%%%%%%

\section{Conclusions}
\label{sec_conclude}

We considered non-logarithmic
heavy-quark mass effects in
the Energy-Energy Correlation (EEC) function in the
two-jet (or back-to-back) limit.
In terms of the usual factorization formula,
such effects are factorized into
the coefficient function (or hard factor)
and into the remainder function only.
This work follows a previous analysis
by two of us \cite{Aglietti:2024zhg} concerning logarithmic
mass effects%
%%%%%%%%%%
\footnote{
By {\it logarithmic mass effects},
we mean {\it leading-twist} logarithmic mass effects, i.e. terms
of the form
$\alpha_S^n \log^k(m^2/Q^2)$, and not
of the form
$\alpha_S^n m^2/Q^2 \log^k\left(m^2/Q^2\right),
\, \alpha_S^n m^4/Q^4 \log^k\left(m^2/Q^2\right),\cdots.$
},
%%%%%%%%%%%
which have been factorized and
resummed to all orders in $\alpha_S$ into a
(massive) Sudakov form factor.

{\color{black} We have defined a partial EEC cumulant by normalizing the EEC
cumulant at a fixed angle $\chi_M<\pi$. This quantity is not a
restricted event sample or a veto on radiation, but a ratio of two
integrals of the same inclusive EEC distribution. At first order, the
back-to-back endpoint coefficient cancels in the ratio, while the forward
endpoint does not enter, so that only the real-emission distribution in
the finite interval between $\chi$ and $\chi_M$ is required. The
calculation can therefore be performed in $D=4$ space-time dimensions
without determining the endpoint contributions separately.}

We have then numerically evaluated the
first-order real corrections, involving
the process
\beq
e^+e^- \to Q \overline{Q} g.
\eeq
We have found it more technically convenient
to evaluate the partially-integrated
EEC distribution rather than the differential
distribution. The former case indeed involves
a simple two-dimensional integral:
the first integration can be easily made analytically,
while the second one is to be made
numerically%
%%%%%%%%%%
\footnote{
In principle, the second integration can 
be made analytically in terms of
generalized harmonic dilogarithms
\cite{Aglietti:2004tq}.
However, the occurrence of nested square
roots involving both $\chi$ and $\eta$
renders the analytic calculation
in real life very hard, if not impossible
(hidden zeroes are expected to occur).
}.
%%%%%%%%%%%
The differential distribution
is then obtained by fitting (interpolating)
the numerical table of the values
of the EEC cumulant and then
differentiating with respect to the
correlation angle $\chi$.
That is to be compared with the direct calculation of the differential distribution, where
the (first) integration over the $\delta$-function
(implementing the EEC kinematic constraint),
produces a numerically rather singular
Jacobian.
A good agreement with previous (numerical) computations of the EEC spectrum
is also found \cite{Csikor:1983dt,Ali:1984gzn}.

By subtracting the first-order expansion
($\mathcal{O}(\alpha_S)$)
of the massive Sudakov form factor
(improved with respect to the one derived in \cite{Aglietti:2024zhg}),
we have evaluated the remainder functions
and the coefficient functions
of the EEC spectra
for a wide range of
values of the mass parameter
$\eta \equiv 2m/Q$ ($0\le \eta \le 1$), where $m$ is the
heavy-quark mass and $Q=\sqrt{s}$ is the
hard scale of the process.

By comparing the {\it ab initio}
massless remainder function (i.e. obtained from
standard factorization of the massless fixed-order
distribution) with the remainder functions
evaluated at very small masses, $\eta \ll 1$,
we have found that the usual factorization
scheme is not smooth in the massless limit.
This unphysical and unpleasant
characteristic is eliminated by introducing
an {\it improved} factorization scheme,
which is continuous in the massless limit
$\eta \to 0^+$, and involves a new
coefficient function which, unlike the
usual one, does depend on the correlation
angle $\chi$.

For ease of reference, let us summarize
the main formulae of the improved
scheme for our partial EEC cumulant:
\bea
R\left( \chi; \, \chi_M ; \, \eta; \, \alpha_S\right)
&\equiv&
\frac{ \int\limits_0^{\chi}
d\Sigma/d\chi'\left( \chi'; \, \eta; \, \alpha_S \right) \, d\chi' }{ \int\limits_0^{\chi_M} d\Sigma/d\chi'\left( \chi'; \, \eta; \, \alpha_S \right) \, d\chi' } \, \simeq
\\ \nonumber
&\simeq&
\left[ 1 \, + \, \frac{C_F\alpha_S}{\pi}
\, \tilde{C}^{(1)}(\chi_M;\eta;\chi)
\right]
\frac{\Phi(\chi;\, \eta,\alpha_S)}{\Phi(\chi_M;\, \eta,\alpha_S)}
\, + \, 
\frac{C_F\alpha_S}{\pi} \,
\mathrm{Rem}^{(1)}_0(\chi);
\eea
{\color{black} where $\chi_M\sim\pi/2$ (more generally, $0<\chi_M<\pi$) is
a fixed normalization angle and $0\leq\chi\leq\chi_M$.  The mass
parameter is $\eta\equiv2m/Q$, with
$Q\equiv\sqrt{s}\gg\Lambda_{QCD}$ and
$\alpha_S\equiv\alpha_S(Q)\ll1$.}
Furthermore:
%%%%%%%%%%%%%%%%%
\begin{enumerate}
%%%%%
\item
$\tilde{C}^{(1)}(\chi_M;\eta;\chi)$
is the improved massive first-order coefficient function, given explicitly by:
\beq
\tilde{C}^{(1)}(\chi_M;\eta;\chi) \, = \,
- \, \mathrm{Rem}^{(1)}_0\left(\chi_M\right)
\, + \, a_\eta \left(
e^{ - \, b_\eta \, \chi}
\, - \, e^{ - \, b_\eta \, \chi_M}
\right).
\eeq
The coefficients $a_\eta$ and
$c_\eta \equiv \eta \, b_\eta $
are given in tables 1 and 2
for the vector and axial cases
respectively;
%%%%%
\item
$\Phi(\chi;\eta,\alpha_S)$ is the massive
Sudakov form factor for the partially-integrated
distribution;
%%%%%
\item
$\mathrm{Rem}^{(1)}_0(\chi)$
is the massless Remainder function,
explicitly given in eq.(\ref{eq_Rem0_analit}).
%%%%%%%%%%%%%%%
\end{enumerate}
A similar discontinuity phenomenon
to the one found here,
has also been found, a few years ago,
in threshold resummation in the massive case;
A similar solution has been found also
in that case \cite{Aglietti:2022rcm,Gaggero:2022hmv,Ghira:2023bxr,Ghira:2024nkk}.
Let us remark that, in general, EEC resummation
in the two-jet limit in the massive
case is a truly multi-scale resummation,
as the heavy quark introduces a genuinely
new mass scale.

{\color{black} Recent studies of heavy-flavour energy correlators, angularities and the
Lund $b$-jet plane provide complementary probes of mass-regulated radiation
and of the dead-cone effect~\cite{Dhani:2024hfs,Dhani:2025lep,Ghira:2025lund,Ghira:2026ang}.
These jet-based observables resolve radiation inside a tagged heavy-flavour
jet, whereas the massive EEC is a global, jet-algorithm-independent
correlator and probes the inclusive recoil generated by soft and
quasi-collinear radiation from the heavy-quark pair.  Measurements of both
classes of observables at different hard scales could test the universality
of perturbative mass effects and help disentangle them from jet-definition
effects and non-perturbative corrections.}

Our work can be extended along several directions of investigation.
Logarithmic heavy-quark mass effects
have been included in our recent
global analysis of EEC data
\cite{Aglietti:2026bhu}.
Having explicitly calculated in this work
the non-logarithmic heavy-quark mass effects considered here,
new phenomenological analyses with improved control over mass effects
are now feasible.

It would be interesting to extend
our results to the next order,
i.e. to evaluate the
coefficient function and the
remainder function to $\mathrm{O}(\alpha_S^2)$,
in particular in the improved scheme,
whenever possible.
The calculation of the massive EEC
spectrum to second order is, of course,
much more complicated
than the first-order one,
and a heavy numerical work, via a Monte Carlo program, is expected.

%%%%%%%%%%%%%%%%%%%%%%%%%%%

%%%%%%%%%

\appendix

\newpage
%%%%%%%%%%%%%%%%%%%%%%%%%%%%%%%%%%%%%%%%%%%%%%
{\color{black}
\section{Three-body kinematics with a massive quark-antiquark pair and a gluon}
\label{sec_kino}
In this appendix we consider the kinematics of the three-body final state 
$e^+e^- \to Q\bar{Q}g$, consisting of a massive quark-antiquark pair and a 
massless gluon. In the Center-Of-Mass (C.O.M.) frame ($\bf{P}_{\mathrm{tot}}=0$), energy conservation is written:
\beq
\label{eq_cons_ener}
x \, + \, \overline{x} \, + \, z \, = \, 2;
\eeq
where we have defined the dimensionless parton energies and mass parameter:
\bea
\label{eqs_def_energ}
x &\equiv& \frac{2 E}{Q};
\nonumber\\
\overline{x} &\equiv& \frac{2 \overline{E}}{Q};
\nonumber\\
z &\equiv& \frac{2 \omega}{Q};
\nonumber\\
\eta &\equiv& \frac{2 m}{Q}.
\eea
$Q \equiv \sqrt{s}$ is the C.O.M. energy,
where $s \equiv \left(l+\overline{l}\right)^2$
is the usual Mandelstam variable,
with $l^\mu$ and $\overline{l}^\mu$
the initial electron and positron
4-momenta respectively.
Finally $E,\overline{E}$ and $\omega$ are the
quark, antiquark and gluon energies
respectively and
$m$ is the heavy quark mass.
Eq.(\ref{eq_cons_ener}) implies that
only two parton energies are independent.

At the level of single parton, the energy ranges
read:
\beq
\eta \, \le \, x \, \le \, 1;
\qquad
\eta \, \le \, \overline{x} \, \le \, 1;
\qquad
0 \, \le \, z \, \le \, 1 \, - \, \eta^2.
\eeq
Note that:
\beq
\eta \, = \, \sqrt{1 \, - \, u^2};
\eeq
where $u\equiv dr/dt$ is the (kinematic)
heavy quark velocity in the soft limit.
General 4-momentum conservation
is written:
\beq
\label{eq_four_mom_con}
q^\mu \, = \, p^\mu \, + \, \overline{p}^\mu \, + \, k^\mu;
\eeq
where:
\beq
q^\mu \, = \, \left( Q ; \, 0,0,0 \right)
\eeq
is the total 4-momentum
of the final system in the C.O.M. frame
(equivalently, the total momentum of the initial $e^+e^-$ pair or the momentum of
the intermediate $\gamma$ or $Z^0$).
$p^\mu$, $\overline{p}^\mu$ and $k^\mu$ are the 4-momenta
of the quark, the antiquark and the gluon
respectively.

Simple kinematic relations are obtained
by taking to the l.h.s., one of the 4-momenta
on the r.h.s. of eq.(\ref{eq_four_mom_con}),
and then squaring at both sides.
In the case of the antiquark momentum,
\beq
q^\mu \, - \, \overline{p}^\mu
\, = \,
p^\mu \, + \, k^\mu,
\eeq
one obtains the equation:
\beq
\label{eq_p_plus_k}
2\left(1 \, - \, \overline{x}\right) \, = \, z
\left(x \, - \, \sqrt{x^2-\eta^2} \cos\theta_{Qg}
\right);
\eeq
where $\theta_{Qg}$ is the angle between
the quark and the gluon space momenta.
A similar relation is obtained by exchanging
quark and antiquark labels
($Q \leftrightarrow \overline{Q}$,
$x \leftrightarrow \overline{x}$):
\beq
\label{eq_pb_plus_k}
2\left(1 \, - \, x\right) \, = \, z
\left(\overline{x} \, - \,
\sqrt{\overline{x}^2 \, - \, \eta^2}
\cos\theta_{\overline{Q}g}
\right).
\eeq
Finally, in the case of the gluon,
\beq
q^\mu \, - \, k^\mu
\, = \,
p^\mu \, + \, \overline{p}^\mu ,
\eeq
one obtains the more complicated equation:
\beq
\label{eq_p_plus_pb}
2\left( 1 \, - \, z \right) \, = \,
\eta^2 \, + \, x \, \overline{x} \, - \,
\sqrt{\left( x^2 \, - \, \eta^2 \right)
\left( \overline{x}^2 \, - \, \eta^2 \right)}
\, \cos\theta_{Q\overline{Q}}.
\eeq
The above eqs.(\ref{eq_p_plus_k}), (\ref{eq_pb_plus_k}) and (\ref{eq_p_plus_pb})
involve all three parton energies and one
relative pair angle.
These equations imply that the energy of
a parton controls the invariant mass
(squared) of the system of the two
recoiling partons.
The larger such energy, the smaller
the invariant mass of the recoiling pair.

Note that, by using energy conservation,
namely eq.(\ref{eq_cons_ener}),
one can express eqs.(\ref{eq_p_plus_k}), (\ref{eq_pb_plus_k}) and (\ref{eq_p_plus_pb})
in terms of two (arbitrarily chosen) independent energies.

Finally, since $Q\overline{Q}g$ final states are planar
(in the COM frame, space momenta add up to zero by definition),
the following geometric relation holds:
\beq
\theta_{Qg} \, + \,
\theta_{\overline{Q}g} \, + \,
\theta_{Q\overline{Q}} \, = \, 2 \pi
\qquad \left(
\theta_{Qg},
\, \theta_{\overline{Q}g}, \, \theta_{Q\overline{Q}}
\, \in \, [0,\pi] \right).
\eeq

%
%%%%%%%%%%%%%%%%%%%
%
\begin{figure}[h]
\begin{center}
\includegraphics[width=0.65\textwidth]{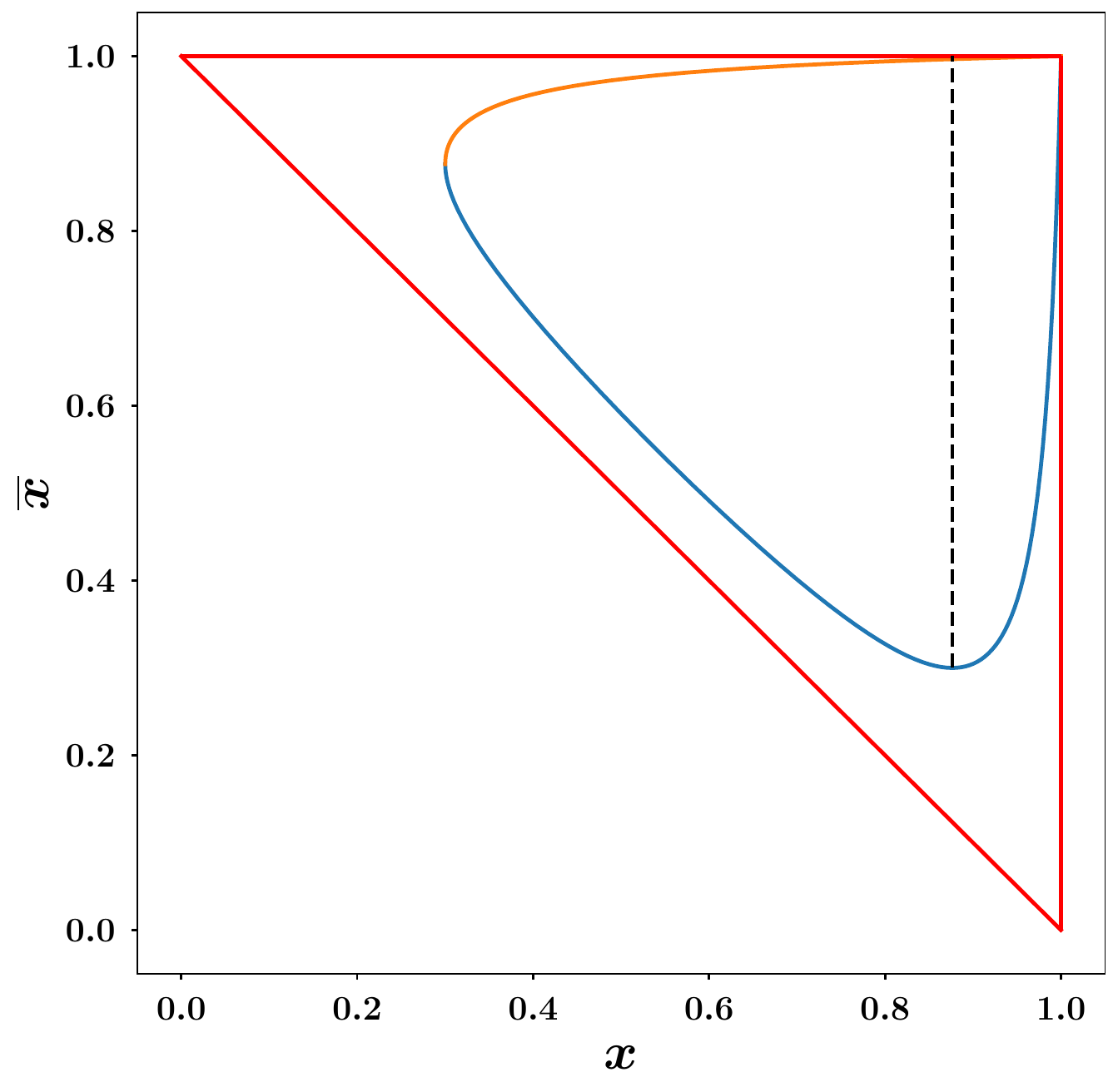}
\footnotesize
\caption{
\label{fig_plotxxbfin}
\it
Allowed region in the quark-antiquark energy plane
(Dalitz plot)
in the massless case (the region
inside the red triangle) and in the massive
case (the region below
the yellow curve $\overline{x}_{\max}=\overline{x}_{\max}(x,\eta)$ and above the blue curve
$\overline{x}_{\min}=\overline{x}_{\min}(x,\eta)$),
for $\eta=0.3$.
The dotted black vertical line,
$x=\tilde{x}(\eta)$
involves the minimal antiquark
energy $\overline{x}_{\min}=\eta$
(see text);
roughly speaking, mass effects
are substantial to the right
of this line.
By increasing $\eta$, the integration
region shrinks.
}
\end{center}
\end{figure}
%
%%%%%%%%%%%
%

%%%%%%%%%%%%%%%%%%%%%%%%%%%%%%%%%%%%%%%%%

\subsection{Quark and antiquark energies}

Since we do not consider ``oriented'' distributions
in ordinary (physical) space
(such as, for example, an oriented shape variable),
it is convenient to use the quark energy $x$ and antiquark energy $\overline{x}$ as independent variables.

In order to find the kinematically-allowed region in the $x$-$\overline{x}$ plane,
one can use, for example, eq.(\ref{eq_p_plus_k}).
In order to eliminate the gluon energy in favor
of the quark and antiquark energies,
we use energy conservation,
eq.(\ref{eq_cons_ener}), in the form
\beq
z \, = \, 2 \, - \, x \, - \, \overline{x}.
\eeq
We then obtain the equation:
\beq
\label{eq_p_plus_k_red}
2\left( 1 \, - \, \overline{x} \right) \, = \,
\left( 2 \, - \, x \, - \, \overline{x} \right)
\left(
x \, - \, \sqrt{x^2 \, - \, \eta^2} \cos\theta_{Qg}
\right).
\eeq
The formula for the antiquark energy as a function of the quark
energy and the quark-gluon angle then reads:
\beq
\label{eq_kino_base}
\overline{x} \, = \,
\overline{x}\left(x,\theta_{Qg}; \eta \right)
\, = \,
2 \, - \, x \, - \,
\frac{2(1 \,- \, x)}{
2 \, - \, x \, + \, \sqrt{x^2 \, - \, \eta^2}\, \cos\theta_{Qg}}.
\eeq
According to the above formula, for any value of $x$ ($\eta$ is fixed in a given physical process),
$\overline{x}$ is a monotonically increasing function of
$\cos\theta_{Qg}$.
Therefore the maximal antiquark energy,
$\overline{x}_{\max}$, is obtained
by setting $\theta_{Qg}=0$.
This fact is quite intuitive: the antiquark
reaches its maximal energy when it recoils
against a collinear quark-gluon pair,
i.e. when $\theta_{Qg}=0$.

With similar reasoning, the minimal antiquark energy
$\overline{x}_{\min}$ is obtained
by setting $\theta_{Qg}=\pi$ in
eq.(\ref{eq_kino_base}).
This latter fact is also rather intuitive: the antiquark reaches its smallest possible energy when the quark
and the gluon are back-to-back,
i.e. $\theta_{Qg}=\pi$, with the antiquark
balancing the space-momenta difference of the
$Qg$ back-to-back pair.
Therefore, by taking $\cos\theta_{Qg}=\mp 1$ respectively, we obtain (see fig.\ref{fig_plotxxbfin}):
\bea
\label{eqs_xmin_and_xmax}
\overline{x}_{\min} &=& \overline{x}_{\min}(x;\eta) \, = \,
2 \, - \, x \, - \, \frac{2(1-x)}{2-x-\sqrt{x^2-\eta^2}};
\nonumber\\
\overline{x}_{\max} &=& \overline{x}_{\max}(x;\eta) \, = \,
2 \, - \, x \, - \,
\frac{2(1 \, - \, x)}{
2 \, - \, x \, + \, \sqrt{x^2 \, - \, \eta^2}}.
\eea
It is immediate to check that the above
formulae have the correct massless limits:
\bea
\lim_{\eta \to 0^+} \overline{x}_{\min}(x;\eta) &=&
1 \, - \, x;
\nonumber\\
\lim_{\eta \to 0^+} \overline{x}_{\max}(x;\eta) &=& 1.
\eea
In the massless limit,
the well-known triangular region
in the $x$-$\overline{x}$ plane is obtained
(see fig.\ref{fig_plotxxbfin}),
with vertices in the points $(1,0)$,
$(0,1)$ and $(1,1)$
(the latter is the ``soft-gluon point'', with the Born kinematics).

The minimum of $\overline{x}_{\min}(x;\eta)$,
as a function of $x$ ($\eta$ is fixed),
i.e. the minimal antiquark energy,
is obtained for:
\beq
\label{eq_minim_quark_ener}
x \, = \, \tilde{x}(\eta);
\eeq
where (see fig.\,\ref{fig_plotxtilde}):
\beq
\label{eq_-def_xtilde}
\tilde{x}(\eta)
\, \equiv \, \frac{2}{2 \, - \, \eta} \, - \, \eta.
\eeq

\begin{figure}[h]
\begin{center}
\includegraphics[width=0.85\textwidth]{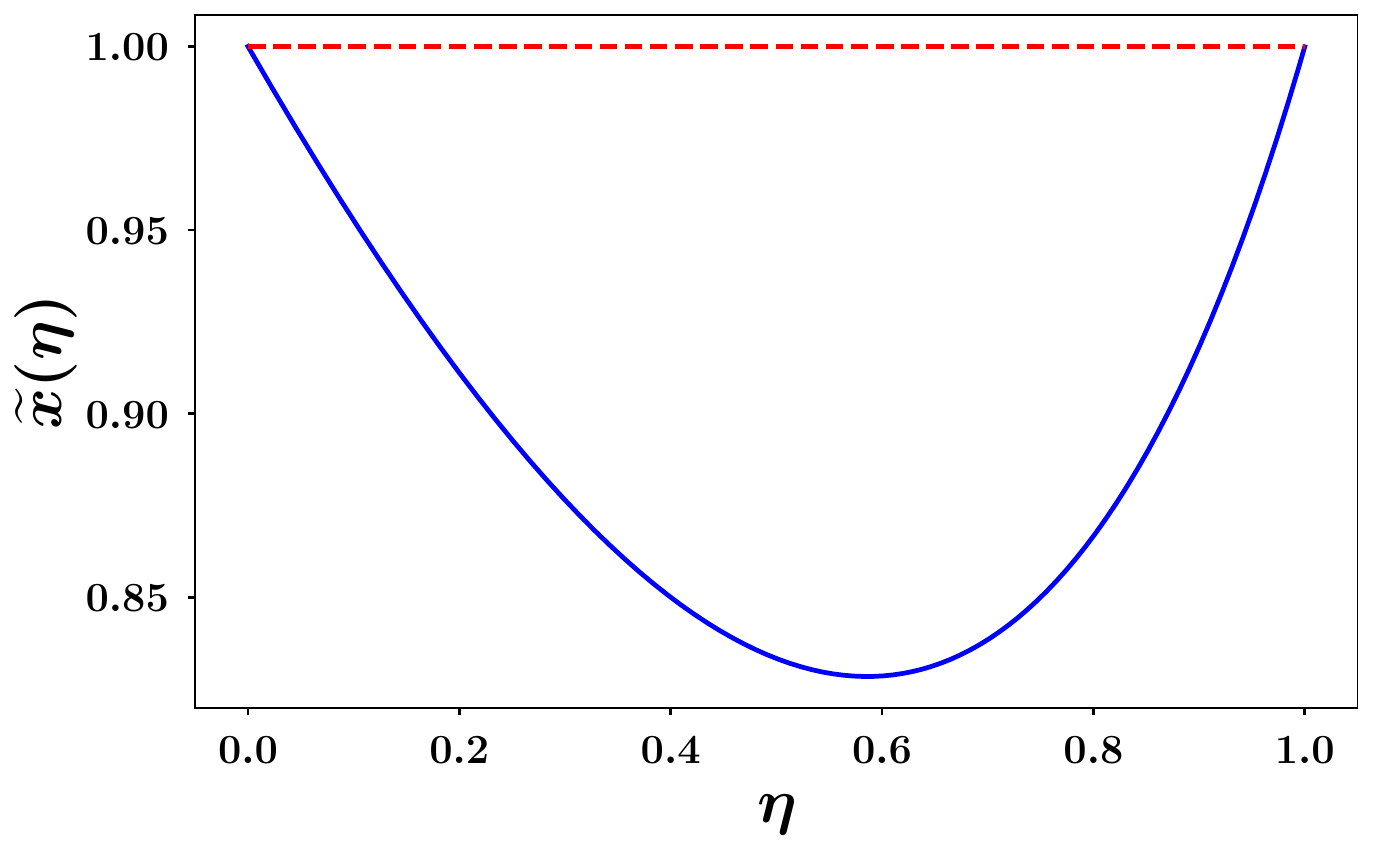}
\footnotesize
\caption{
\label{fig_plotxtilde}
\it
$\tilde{x}=\tilde{x}(\eta)$, i.e. as a function of the mass parameter $\eta \in [0,1]$.
}
\end{center}
\end{figure}

\noindent
The minimal antiquark energy equals its mass,
as it should:
\beq
\overline{x}_{\min}
\left[ x = \tilde{x}(\eta);
\,\eta\right] \, = \, \eta.
\eeq
The value of $x$ in
eq.(\ref{eq_minim_quark_ener}),
corresponds
to the kinematic configuration
with the antiquark at rest,
and the quark and the gluon
back-to-back sharing the available
energy $Q-m$.

Since the ``external'' variable $x$ has the range
$\eta \, \le \, x \, \le \, 1$,
in order to
evaluate a generic distribution,
we have to compute a double integral of the form:
\beq
\int\limits_\eta^1 dx
\int\limits_{\overline{x}_{\min}(x;\eta)}^{\overline{x}_{\max}(x;\eta)}
d \overline{x} \, F\left(x,\overline{x}; \chi, \eta \right);
\eeq
where the function $F\left(x,\overline{x}; \chi, \eta \right)$
contains the $Q\overline{Q}g$ matrix elements
as well as the kinematical constraints
defining the distribution.

%
%%%%%%%%%%%%%%%%%%%
%

\section{Matrix elements with a massive quark-antiquark pair and a gluon}
\label{sec_ME}

The double differential cross section
in the quark and antiquark energies
$x$ and $\overline{x}$
for
\beq
e^+ e^- \, \to \, Q \overline{Q} g,
\eeq
can be decomposed as
\cite{Dokshitzer:1994jt}:
\beq
\frac{d^2\sigma}{dx d\overline{x}}
\, = \, \frac{d^2\sigma_{VV}}{dx d\overline{x}}
\, + \,
\frac{d^2\sigma_{AA}}{dx d\overline{x}}.
\eeq
There is no interference term
involving both the vector $(V)$ and
the axial $(A)$ current, i.e. a term
\beq
\frac{d^2\sigma_{VA}}{dx d\overline{x}},
\eeq
because we have integrated
over the orientation of the
plane containing the $Q\overline{Q}g$ system
with respect to the straight line identified
by the initial $e^+e^-$ pair
(we work in the C.O.M. system).

The vector and axial contributions
to the distribution are written:
\beq
\frac{1}{\sigma_{CC}^{(0)}}
\, \frac{d^2\sigma_{CC}}{dx d\overline{x}}
\, = \,
\frac{C_F \alpha_S}{2\pi} \,
\mathcal{M}_{CC}\left(x,\overline{x}; \, \eta \right);
\qquad C = V, A;
\eeq
where $C_F=(N^2-1)/(2N)=4/3$
for $N=3$ colors in QCD.

The lowest-order vector and axial contributions to the total
cross section read:
\bea
\sigma_{VV}^{(0)}
&=&
v \frac{3-v^2}{2}
\left[
e_Q^2 \frac{\pi \alpha^2}{s}
- 4\pi k \alpha \, V_e \, e_Q \,  V_Q
\frac{s-m_Z^2}{ (s-m_Z^2)^2 + (s \Gamma_Z/m_Z)^2 } \, +
\right.
\nonumber\\
&& \left. \qquad\quad\,\, + \,
4\pi k^2
\left( V_e^2 + A_e^2 \right) V_Q^2
\frac{s}{ (s-m_Z^2)^2 + (s \Gamma_Z/m_Z)^2 }
\right];
\nonumber\\
\sigma_{AA}^{(0)} &=&
v^3 \, 4 \pi k^2
\left( V_e^2 + A_e^2 \right) A_Q^2
\frac{s}{ (s-m_Z^2)^2 + (s \Gamma_Z/m_Z)^2 };
\eea
where:%
%%%%%%%%%%
\footnote{
The constant $k$ in \cite{Dokshitzer:1994jt}
contains an additional factor $\alpha$ at the
denominator.
}
%%%%%%%%%
\beq
k \, \equiv \, \frac{\sqrt{2} G_F m_Z^2 }{4\pi}.
\eeq
$V_e$ and $A_e$ are
the vector and the axial couplings
to the $Z^0$ of the electron
respectively.
Finally, $e_Q, V_Q$ and $A_Q$ are the electric charge,
the vector and the axial couplings
to the $Z^0$ of the heavy quark $Q$
respectively.

%%%%%%%%%%%%%%%%%%%%%%%%%%%

\subsection{Vector current}

In the case of a vector current:
\bea
\label{eq_ME_VV}
\mathcal{M}_{VV}\left(x,\overline{x};\eta \right)
&\equiv&
\frac{1}{ \sqrt{1 \, - \, \eta^2} }
\Bigg\{
\,\, \frac{ 2x + 2\overline{x} - 2 - \eta^2  }{(1-x)(1-\overline{x})}
\, - \, \frac{\eta^2}{2}
\left[
\frac{ 1 }{(1-x)^2}
\, + \,
\frac{ 1 }{(1-\overline{x})^2}
\right] \, +
\nonumber\\
&& \quad\qquad\qquad\qquad\qquad\qquad\qquad
+ \, \frac{1}{1+\eta^2/2} \,
\frac{ (1-x)^2 + (1-\overline{x})^2 }{(1-x)(1-\overline{x})}
\Bigg\}
\, =
\nonumber\\
&=& \frac{1}{ \sqrt{1 \, - \, \eta^2} }
\Bigg\{
\frac{ x^2 + \overline{x}^2 - \eta^2  }{(1-x)(1-\overline{x})}
\, - \, \frac{\eta^2}{2}
\left[
\frac{ 1 }{(1-x)^2}
\, + \,
\frac{ 1 }{(1-\overline{x})^2}
\right] \, +
\nonumber\\
&& \quad\qquad\qquad\qquad\qquad\qquad\qquad
\, - \, \frac{\eta^2}{2+\eta^2} \,
\frac{ (1-x)^2 + (1-\overline{x})^2 }{(1-x)(1-\overline{x})}
\Bigg\}.
\eea
The following remarks are in order:
%%%%%%%%%%%%%%%%%
\begin{enumerate}
%%%%%
\item
Two equivalent representations of the matrix
element have been given above, the second one
having a simple
massless limit $\eta \to 0^+$.
In this limit, indeed,
the second and the third term in the curly bracket
above vanish and one recovers
the well-known massless distribution:
\beq
\label{eq_double_dist_no_mass}
\frac{1}{\sigma_0} \, \frac{d^2\sigma_0}{dx \, d \overline{x}}
\, = \,
\frac{C_F \alpha_S}{2\pi} \,
\frac{ x^2 + \overline{x}^2 }{(1-x)(1-\overline{x})}.
\eeq
The subscript ``$0$'' here denotes massless quantities
(and not tree-level ones, as usual).
Note that the corrections to the massless limit
are $\mathcal{O}(\eta^2)$ i.e.,
as expected, {\it quadratic}
in the quark mass;
%%%%%
\item
The second term in the curly brackets above
(involving the square bracket containing two squares),
vanishing in the massless limit,
is related to self-energy type
diagrams in Feynman gauge;
%%%%%
\item
In the most infrared-singular term,
namely the first one,
we can split the soft-singular terms
from the hard quasi-collinear ones, by writing:
\beq
\frac{ x^2 + \overline{x}^2 - \eta^2 }{(1-x)(1-\overline{x})}
\, = \,
\frac{ 2  - \eta^2 }{(1-x)(1-\overline{x})}
\, - \, \frac{ 1 + \overline{x} }{1 - x }
\, - \, \frac{ 1 + x }{1 - \overline{x}}.
\eeq
In the soft limit, $z \to 0^+$,
we do not assume $\eta$
to be small, but rather to be a generic
$\mathcal{O}(1)$
quantity, while in the quasi-collinear
limit we do assume $\eta \ll 1$
(otherwise there is no collinear enhancement).
%%%%%
\item
The last term can be written
in different ways, as:
\beq
\frac{ (1-x)^2 + (1-\overline{x})^2 }{(1-x)(1-\overline{x})}
= \frac{ 1-x }{1-\overline{x}}
+\frac{1-\overline{x} }{1-x}
= \frac{ z^2 }{(1-x)(1-\overline{x})} - 2.
\eeq
%%%%%%%%%%%%%%%
\end{enumerate}

%%%%%%%%%%%%%%%%%%%%%%%%%%

\subsection{Axial current}

In the case of an axial current, the
matrix-element squared for
$e^+e^-\mapsto Q \overline{Q} g$ reads \cite{Dokshitzer:1994jt}:
\bea
\label{eq_ME_AA}
\mathcal{M}_{AA}\left(x,\overline{x};\eta \right)
&\equiv&
\frac{1}{ \sqrt{1 \, - \, \eta^2} }
\Bigg\{
\,\, \frac{ 2x + 2\overline{x} - 2 - \eta^2  }{(1-x)(1-\overline{x})}
\, - \, \frac{\eta^2}{2}
\left[
\frac{ 1 }{(1-x)^2}
\, + \,
\frac{ 1 }{(1-\overline{x})^2}
\right]
\, +
\nonumber\\
&& \qquad\qquad\qquad\qquad\,\,\,
+ \, \frac{1 \, + \, \eta^2/2}{1 \, - \, \eta^2} \,
\frac{ (1-x)^2 + (1-\overline{x})^2 }{(1-x)(1-\overline{x})}
\, + \, \frac{\eta^2}{1-\eta^2}
\Bigg\}
\, =
\nonumber\\
&=& \frac{1}{ \sqrt{1 \, - \, \eta^2} }
\Bigg\{
\frac{ x^2 + \overline{x}^2 - \eta^2  }{(1-x)(1-\overline{x})}
\, - \, \frac{\eta^2}{2}
\left[
\frac{1}{(1-x)^2}
\, + \,
\frac{1}{(1-\overline{x})^2}
\right] \, +
\nonumber\\
&& \qquad\qquad\qquad\qquad
\, + \, \frac{3}{2}\,\frac{\eta^2}{1-\eta^2} \,
\frac{ (1-x)^2 + (1-\overline{x})^2 }{(1-x)(1-\overline{x})}
\, + \, \frac{\eta^2}{1-\eta^2}
\Bigg\}.
\eea
The following remarks are in order:
\begin{enumerate}
%%%%%%%%%%%%%%%%%
\item
%%%%%
The distribution above has the
correct massless limit, eq.(\ref{eq_double_dist_no_mass}),
and differences between the
vector and the axial case are
$\mathcal{O}(\eta^2)$.
\item
The first two terms in the curly bracket, on the first line
--- the most singular one, $\propto 1/(1-x)/(1-\overline{x})$, and the self-energies related one ---,
are exactly the same for the two currents.
\item
The differences with respect to the vector case
are:
%%%%%%%%%%%%%%%%%
\begin{enumerate}
%%%%%
\item
A different coefficient in the third
term in the curly bracket
--- which is positive rather than negative.
%%%%%
\item
The occurrence
of a constant term, independent of $x$
and $\overline{x}$ --- the last one.
%%%%%%%%%%%%%%%
\end{enumerate}
%%%%%%%%%%%%%%%
\end{enumerate}
}
\end{document}